%% file: main.tex
\pdfoutput=1
\newcommand*{\ATLASLATEXPATH}{latex/}
\documentclass[cernpreprint,texlive=2011,txfonts,UKenglish]{\ATLASLATEXPATH atlasdoc}

\usepackage[biblatex=false]{\ATLASLATEXPATH atlaspackage}
\usepackage{\ATLASLATEXPATH atlasbiblatex}

\usepackage{\ATLASLATEXPATH atlascontribute}

\usepackage[other=true]{\ATLASLATEXPATH atlasphysics}

\graphicspath{{logos/}{figures/}}

\usepackage{mathrsfs}
\usepackage{rotating}
\usepackage{a4wide}

\newboolean{Nothing}
\setboolean{Nothing}{true}
\newboolean{Internal}
\setboolean{Internal}{true}

\newboolean{BdFraction3.3}
\setboolean{BdFraction3.3}{true}
\input{keys}

\input{main-metadata}
\hypersetup{pdftitle={ATLAS document},pdfauthor={The ATLAS Collaboration}}

\begin{document}

\maketitle

\input{Sections1-2}

\input{Section3}

\input{Section4}

\input{Section5}

\input{Section6}

\input{Section7}

\input{Sections8-9}

\section*{Acknowledgements}

\input{acknowledgements/Acknowledgements}

\bibliographystyle{bibtex/bst/atlasBibStyleWithTitle}
\bibliography{atlaspaper.bib}

\newpage \input{atlas_authlist}

\end{document}

%% file: main-metadata.tex
\AtlasTitle{Measurement of the CP-violating phase $\phis$ and the $\Bs$ meson decay width difference with $\Bst$ decays in ATLAS}

\author{The ATLAS Collaboration}

\PreprintIdNumber{CERN-PH-2015-166}

\AtlasJournalRef{JHEP 08 (2016) 147}
\AtlasDOI{10.1007/JHEP08(2016)147}

\AtlasAbstract{%
 A measurement of the $B^0_s$ decay parameters in the \Bst\ channel using  an integrated luminosity of \ilumi\ collected by the ATLAS detector from \CoMEnergy\ $pp$ collisions at the LHC is presented. The measured parameters include the $CP$-violating phase $\phi_s$, the decay width $\Gamma_s$ and the width difference between the mass eigenstates $\Delta \Gamma_s$. The values measured for the physical parameters are statistically combined with those from 4.9~$\rm{fb}^{-1}$ of \CoMEnergySeven\ data, leading to the following:
\begin{eqnarray*}
\phis & = & \pSfitCombi \pm \pSstatCombi~\rm{(stat.)}\pm \pSsystCombi~\rm{(syst.)~rad}  \\
\DGs & = & \DGfitCombi \pm \DGstatCombi~\rm{(stat.)}\pm \DGsystCombi~\rm{(syst.)~ps}^{-1} \\
\Gs & = & \GSfitCombi \pm \GSstatCombi ~\rm{(stat.)}\pm \GSsystCombi ~\rm{(syst.)~ps}^{-1}.
\end{eqnarray*}
In the analysis  the parameter \DGs ~ is constrained to be  positive.  Results for \phis\ and \DGs\ are also presented as 68\% and  95\% likelihood contours in the \phis--\DGs\  plane. Also measured in this decay channel are the transversity amplitudes and corresponding strong phases. All measurements are in agreement with the Standard Model predictions.
}

%% file: Sections1-2.tex
\section{Introduction}
New phenomena beyond the predictions of the Standard Model (SM) may alter $CP$ violation in $b$-hadron decays. A channel that  is expected to be sensitive to  new physics contributions is the decay \Bst. $CP$ violation in the \Bst\ decay occurs due to interference between direct decays and decays with \BsaBs\ mixing.  The oscillation frequency of \Bs\ meson mixing is characterized by the mass difference \dms\ of the heavy (\BH) and light (\BL)  mass eigenstates. The $CP$ violating phase \phis\ is defined as the weak phase difference between the \BsaBs\ mixing amplitude and the $b \rightarrow c \overline{c} s$ decay amplitude.  In the absence of $CP$ violation, the \BH\ state would correspond to the $CP$-odd state and the \BL\ to the $CP$-even state.  In the SM the phase \phis\ is small and can be related to Cabibbo--Kobayashi--Maskawa (CKM) quark mixing matrix elements via the relation $\phis \simeq -2 \beta_{s}$, with $\beta_{s} = \mathrm{arg} [- (V_{ts} V^{*}_{tb})/(V_{cs} V_{cb}^{*}) ]$; assuming no physics beyond the SM contributions to \Bs\ mixing and decays, a value of  $-2 \beta_{s} = -0.0363^{+0.0016}_{-0.0015} $ rad can be predicted by combining beauty and kaon physics observables \cite{PhysRevD.84.033005}.

Other physical quantities involved in \BsaBs\ mixing are the decay width $\Gs=(\GL + \GH)/2$ and the width difference $\DGs = \GL - \GH$, where $\GL$ and $\GH$ are the decay widths of the different eigenstates. The width difference is predicted to be $\DGs = 0.087 \pm 0.021$ ps$^{-1}$ \cite{DeltaGamma}.  Physics beyond the SM is not expected to affect \DGs\ as significantly as \phis~\cite{Nierste}. However, extracting \DGs\ from data is interesting as it allows theoretical predictions to be tested~\cite{Nierste}. Previous measurements of these quantities have been reported by the D\O , CDF,  LHCb, ATLAS and CMS collaborations \cite{Abazov:2011ry, CDF:2012ie, LHCb:2013oba, LHCb:2014run1, tagCPV2011, CMSPaper}.

The decay of the pseudoscalar \Bs\ to the vector--vector $J/\psi(\mumu)\phi(K^+K^-)$ final state results in an admixture of $CP$-odd and $CP$-even states, with orbital angular momentum $L = 0$, 1 or 2. The final states with orbital angular momentum $L = 0$ or 2 are $CP$-even,  while the state with $L = 1$ is $CP$-odd.   The same final state can also be produced with $K^{+}K^{-}$ pairs in an $S$-wave configuration \cite{StoneZhang}. This $S$-wave final state is $CP$-odd.  The $CP$ states are separated statistically using an angular analysis of the final-state particles. Flavour tagging is used to distinguish between the initial \Bs\ and \aBs\ states.

The analysis presented here  provides a measurement of the \Bst\ decay parameters using \ilumi\ of LHC {\it pp} data collected by the ATLAS detector during 2012 at a centre-of-mass energy of \CoMEnergy. This is an update of the previous  flavour-tagged time-dependent angular analysis of  $\Bs \rightarrow J/\psi \phi$   \cite{tagCPV2011} that was performed using  4.9~$\rm{fb}^{-1}$ of data collected at \CoMEnergySeven. Electrons are now included, in addition to final-state muons, for the flavour tagging using leptons.

\section{ATLAS detector and Monte Carlo simulation} \label{sec:sample}
The ATLAS detector \cite{ATLAS} is a multi-purpose particle physics detector with a forward-backward symmetric cylindrical geometry and nearly $4\pi$ coverage in solid angle.\footnote{ATLAS uses a right-handed coordinate system with its origin at the nominal interaction point (IP) in the centre of the detector and the $z$-axis along the beam pipe. The $x$-axis points from the IP to the centre of the LHC ring, and the $y$-axis points upward. Cylindrical coordinates $(r,\phi)$ are used in the transverse plane, $\phi$ being the azimuthal angle around the beam pipe. The pseudorapidity is defined in terms of the polar angle $\theta$ as $\eta=-\ln\tan(\theta/2)$.} The inner tracking detector (ID) consists of a silicon pixel detector, a silicon microstrip detector  and a transition radiation tracker. The ID is surrounded by a thin superconducting solenoid providing a 2\,T axial magnetic field, and by a high-granularity liquid-argon (LAr) sampling electromagnetic calorimeter. A steel/scintillator tile calorimeter provides hadronic coverage in the central rapidity range. The end-cap and forward regions are instrumented with LAr calorimeters for electromagnetic and hadronic measurements. The muon spectrometer (MS) surrounds the calorimeters and consists of three large superconducting toroids with eight coils each, a system of tracking chambers, and detectors for triggering.

The muon and tracking systems are of particular importance in the reconstruction of \B meson candidates. Only data collected when both these systems were operating correctly and when the LHC beams were declared to be stable are used in the analysis.  The data were collected during a period of rising instantaneous luminosity, and the trigger conditions varied over this time.  The triggers used to select events for this analysis are based on identification of a \Jpsitomm decay, with transverse momentum (\pT)
thresholds of either  $4$ \GeV\ or $6$ \GeV\ for the muons.
The measurement uses \ilumi\ of $pp$ collision data collected with the ATLAS detector at a centre-of-mass energy of \CoMEnergy.  Data collected at the beginning of the  \CoMEnergy \ data-taking period are not included in the analysis due to a problem with the trigger tracking algorithm. The trigger was subsequently changed to use a different tracking algorithm that did not have this problem.

To study the detector response, estimate backgrounds and model systematic effects, 12 million Monte Carlo (MC) simulated \Bst\ events were generated using  {\sc Pythia} 8 \cite{Pythia, Sjostrand:2007gs} tuned with  ATLAS data~\cite{ATL-PHYS-PUB-2011-009}. No \pT\  cuts were applied at the generator level.  The detector response was simulated using the ATLAS simulation framework based on GEANT4~\cite{Atlfast, Geant4}.  In order to take into account the varying  number of  proton--proton interactions per bunch crossing (pile-up) and trigger configurations during data-taking, the MC events were weighted to reproduce the same pile-up and trigger conditions in data.  Additional samples of the  background decay $\Bd \ra J/\psi K^{0*}$, as well as the more general $b\bar{b} \ra J/\psi X$ and $pp \ra J/\psi X$ backgrounds were also simulated using  {\sc Pythia} 8.

%% file: Section3.tex
\section{Reconstruction and candidate selection}
Events must pass the trigger selections described in Section~\ref{sec:sample}. In  addition, each event must contain at least one reconstructed primary vertex, formed from at least four ID tracks, and at least one pair of oppositely charged muon candidates that are reconstructed using information from the MS and the ID \cite{Aad20119}.   A muon identified using a combination of MS and ID track parameters is referred to as a {\it combined-muon}. A muon formed from a MS track segment that is not associated with a MS track but is matched to an ID track extrapolated to the MS is referred to as a {\it  segment-tagged muon}. The muon track parameters are determined from the ID measurement alone, since the precision of  the measured track parameters is dominated by the ID track reconstruction in the  \pT\ range of interest for this analysis. Pairs of oppositely charged muon tracks are refitted to a common vertex and the pair is accepted for further consideration if the quality of the fit meets the requirement \cutChiJpsi. The invariant mass of the muon pair is calculated from the refitted track parameters.  In order to account for  varying  mass resolution in different parts of the detector, the \Jpsi\ candidates are divided into three subsets  according to the pseudorapidity $\eta$ of the muons. A maximum-likelihood fit is used to extract the $J/\psi$ mass and the corresponding mass resolution for these three subsets.  When both muons have $|\eta| <  1.05$, the dimuon invariant mass must fall in the range \cutMassJpsiBB \ to be accepted as a  \Jpsi\ candidate. When one muon has $1.05  < |\eta| <  2.5 $ and the other muon  $|\eta| <  1.05$, the corresponding signal region is \cutMassJpsiEB. For the third subset, where both muons have $1.05  < |\eta| <  2.5 $, the signal region is \cutMassJpsiEE .  In each case the signal region is defined so as to  retain 99.8\% of the \Jpsi\ candidates identified in the fits.  

The candidates for  the decay \phiKK are reconstructed from all pairs of oppositely charged particles with $\pT > 1$~\GeV \ and $|\eta| < 2.5$ that are not identified as muons. Candidate events for \Bsto\ decays are selected by fitting the tracks for each combination of \Jpsitomm\ and \phiKK to a common vertex.  Each of the four tracks is required to have at least one hit in the pixel detector and at least four hits in the silicon microstrip detector. The fit is further constrained by fixing the invariant mass calculated from the two muon tracks to the \Jpsipa mass~\cite{PDG}. A quadruplet of tracks is accepted for further analysis if the vertex fit has a  \cutChiBs, the fitted  \pT\ of each track from \phiKK\ is greater than 1 GeV and the invariant mass of the track pairs (assuming that they are kaons) falls within the interval \cutMassPhi. If there is more than one accepted candidate in the event, the candidate with the lowest $\chi^2 /$d.o.f. is selected.  In total, \ntotaltext \ \Bs\ candidates are collected within a mass range of \cutMassBsRange.

For each \Bs\ meson candidate the proper decay time $t$ is estimated using the expression:
\begin{equation*}
  t = \frac{ L_{xy}\ m_{{}_{B}} }     { p_{ \mathrm{T}_{B}  }}, 
\end{equation*}
where $p_{ \mathrm{T}_{B}  }$ is the  reconstructed transverse momentum of the \Bs\ meson candidate and  $m_{{}_{B}}$ denotes  the  mass  of the \Bs\ meson, taken from  \cite{PDG}. The transverse decay length, $L_{xy}$, is the displacement in the transverse plane of the  \Bs\ meson  decay vertex with respect to the primary vertex, projected onto the direction of the \Bs\ transverse momentum. The position of the primary vertex used to calculate this quantity is determined from a refit following the removal of the tracks used to reconstruct the \Bs\ meson candidate.

For the selected events the average number of pile-up proton--proton interactions is 21, necessitating a choice of the best candidate for the primary vertex at which the \Bs\  meson is produced. 
The variable used is the three-dimensional impact parameter $d_0$, which is calculated as the distance between the line extrapolated from the reconstructed  \Bs\  meson vertex in the direction of the  \Bs\  momentum, and each primary vertex candidate. The chosen primary vertex is the one with the smallest $d_0$.

A study~\cite{ATL-PHYS-PUB-2013-010}  made using a MC simulated dataset has shown that the precision of  the reconstructed \Bs\  proper decay time remains stable over the range of pile-up encountered during 2012 data-taking. 
No \Bs\ meson decay-time cut is applied in this analysis.

%% file: Section4.tex
\section{Flavour tagging}\label{sec:Tagging}
The initial flavour of a neutral $B$ meson can be inferred using information from the opposite-side $B$ meson that contains the other pair-produced $b$-quark in the event~\cite{Field:1977fa, ATLAS-CONF-2013-086}.  
This is referred to as opposite-side tagging (OST).

To study and calibrate the OST methods, events containing $B^{\pm}\to J/\psi K^{\pm}$ decays are used, where the flavour of the $B^{\pm}$-meson  is provided by the kaon charge. A sample of $B^{\pm}\to J/\psi K^{\pm}$ candidates is selected from the entire 2012 dataset satisfying the data-quality selection described in Section 2. Since the OST calibration is not affected by the trigger problem at the start of the 8 \TeV\ data-taking period, the tagging measurement  uses $ 19.5\, \rm fb^{-1} $\ of integrated luminosity of $pp$ collision data. 

\subsection{$B^{\pm}\to J/\psi K^{\pm}$ event selection}

In order to select candidate $B^{\pm}\to J/\psi K^{\pm}$ decays, firstly $J/\psi$ candidates are selected from pairs of oppositely charged combined-muons forming a good vertex, following the criteria described in Section 3.  Each muon is required to have a transverse momentum of at least $4$ \GeV\ and pseudorapidity within $|\eta|<2.5$. The invariant mass of the dimuon candidate is required to satisfy $2.8$ \GeV\ $< m(\mu^+\mu^-) < 3.4$ \GeV. To form the $B$ candidate, an additional track, satisfying the same quality requirements described for tracks in Section 3, is combined with the dimuon candidate using the charged kaon mass hypothesis, and a vertex fit is performed with the mass of the dimuon pair constrained to the known value of the $J/\psi$ mass. To reduce the prompt component of the combinatorial background, a requirement is applied to the transverse decay length of the $B$ candidate of $L_{xy}>0.1$~mm.

A sideband subtraction method is used in order to study parameter distributions corresponding to the $B^{\pm}$ signal processes with the background component subtracted. Events are divided into sub-sets into five intervals in the pseudorapidity of the $B$ candidate and three mass regions.  The mass regions are defined as a signal region around the fitted peak signal mass position $\mu \pm 2\sigma$ and the sideband regions are defined as  $[\mu - 5\sigma,\mu-3\sigma]$ and $[\mu+3\sigma, \mu+5\sigma]$, where $\mu$ and $\sigma$ are the mean and width of the Gaussian function describing the $B$ signal mass.  Separate binned extended maximum-likelihood fits are performed to the invariant mass distribution in each region of pseudorapidity.

\begin{figure}[t]
\begin{center}

\ifthenelse {\boolean{Nothing}}
{\includegraphics[width=0.60\textwidth]{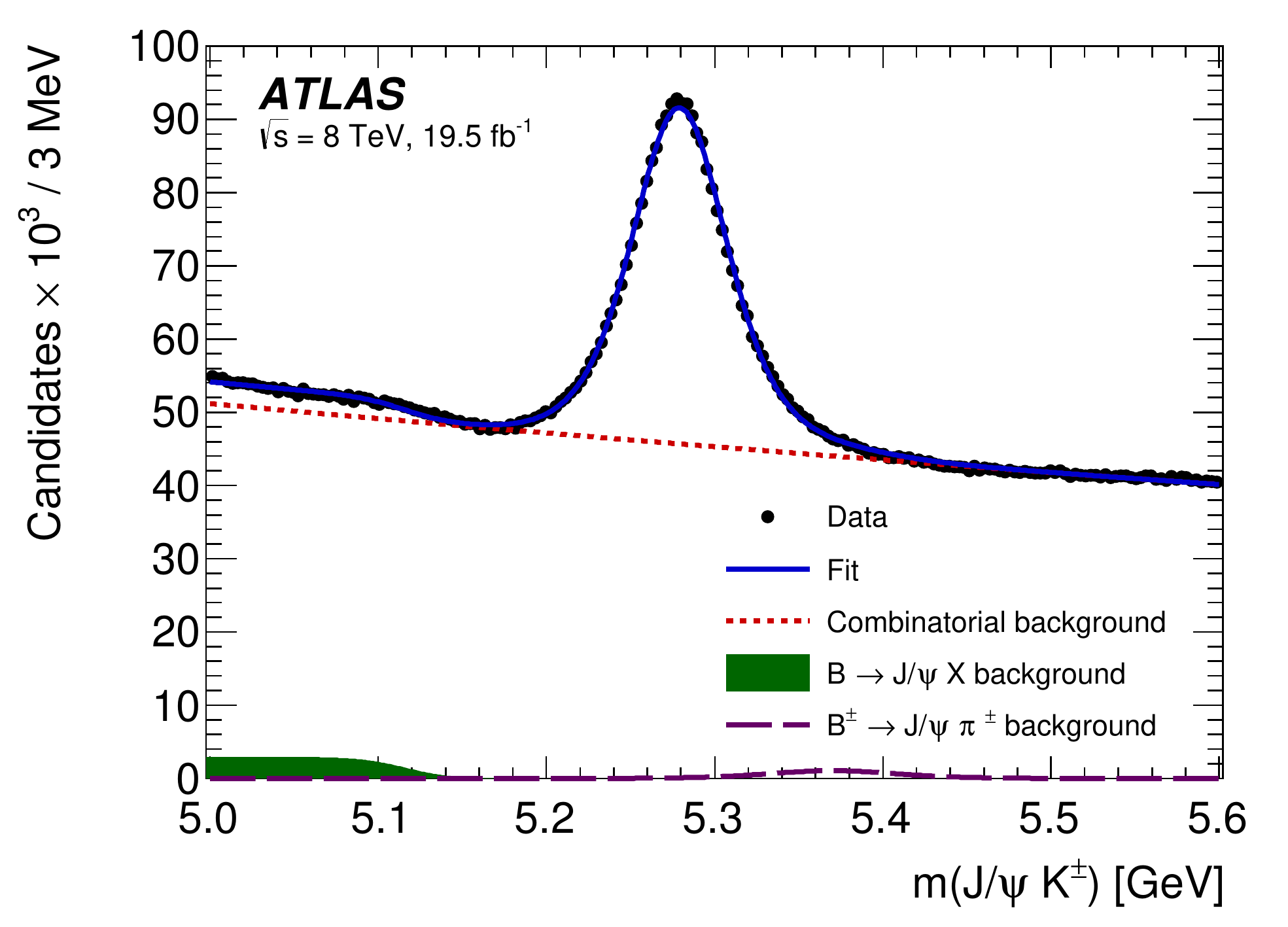}}
{
\ifthenelse {\boolean{Internal}}
{\includegraphics[width=0.60\textwidth]{bplus_massplot_Internal.pdf}}
{\includegraphics[width=0.60\textwidth]{bplus_massplot_Preliminary.pdf}}
}
\caption{The invariant mass distribution for  $B^{\pm}\to\jpsi K^{\pm}$ candidates satisfying the selection criteria, used to study the flavour tagging.
Data are shown as points, and the overall result of the fit is given by the blue curve. 
The contribution from the combinatorial background component is indicated by the red dotted line,  partially reconstructed $B$ decays by the green shaded area, and decays of $B^{\pm}\to J/\psi \pi^{\pm}$, where the pion is mis-assigned a kaon mass, by the purple dashed line.
\label{fig:bplusmass}}
\end{center}
\end{figure}

An exponential function is used to model the combinatorial background and a hyperbolic tangent function to parameterize the low-mass contribution from incorrectly or partially reconstructed $B$ decays. A Gaussian function is used to model the $B^{\pm}\to J/\psi \pi^{\pm}$ contribution. The contribution from non-combinatorial background is found to have a negligible effect on the tagging procedure. Figure~\ref{fig:bplusmass} shows the invariant mass distribution of $B$ candidates for all rapidity regions overlaid with the fit result for the combined data.

\subsection{Flavour tagging methods}\label{sec:FTMethods}
Several methods that differ in efficiency and discriminating power are available to infer the flavour of the opposite-side $b$-quark. The measured charge of a muon or electron from a semileptonic decay of the $B$ meson provides strong separation power; however, the $b\to \ell$ transitions are diluted through neutral $B$ meson oscillations, as well as by cascade decays $b\to c\to \ell$, which can alter the charge of the lepton relative to those from direct $b \to \ell$ decays.  The separation power of lepton tagging is enhanced by considering a weighted sum of the charge of the tracks in a cone around the lepton, where the weighting function is determined separately for each tagging method by optimizing the tagging performance. If no lepton is present, a weighted sum of the charge of tracks in a jet associated with the opposite-side $B$ meson decay provides some separation. The flavour tagging methods are described in detail below.

For muon-based tagging, an additional muon is required in the event, with $\pT>2.5$ \GeV, $|\eta|<2.5$ and with $|\Delta z|<5$~mm from the primary vertex. Muons are classified according to their reconstruction class, {\it combined} or {\it segment-tagged}, and subsequently treated as distinct flavour tagging methods. In the case of multiple muons, the muon with the highest transverse momentum is selected.

\begin{figure}[t]
\begin{center}

\ifthenelse {\boolean{Nothing}}
{\includegraphics[width=0.45\textwidth]{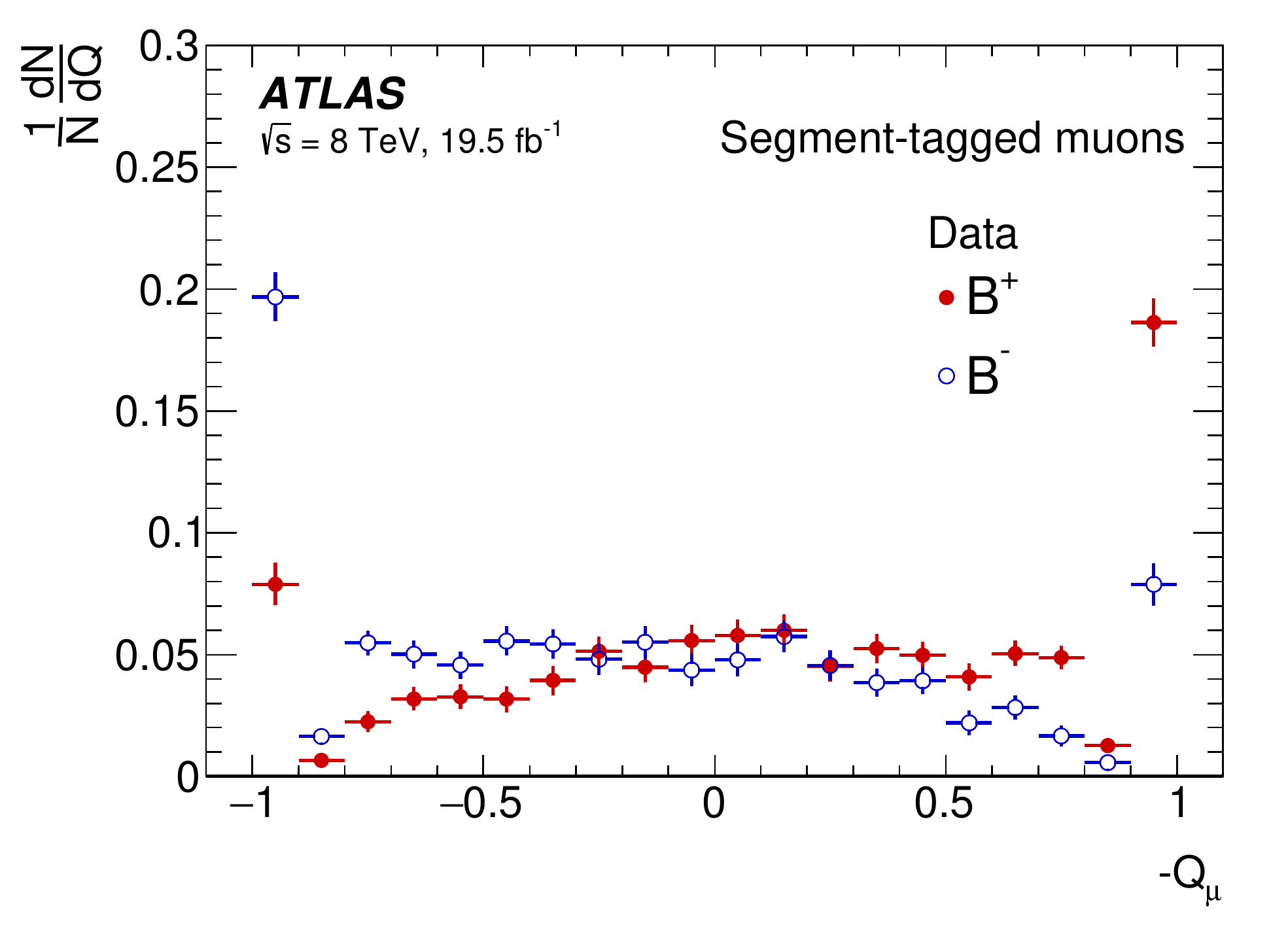}
\includegraphics[width=0.45\textwidth]{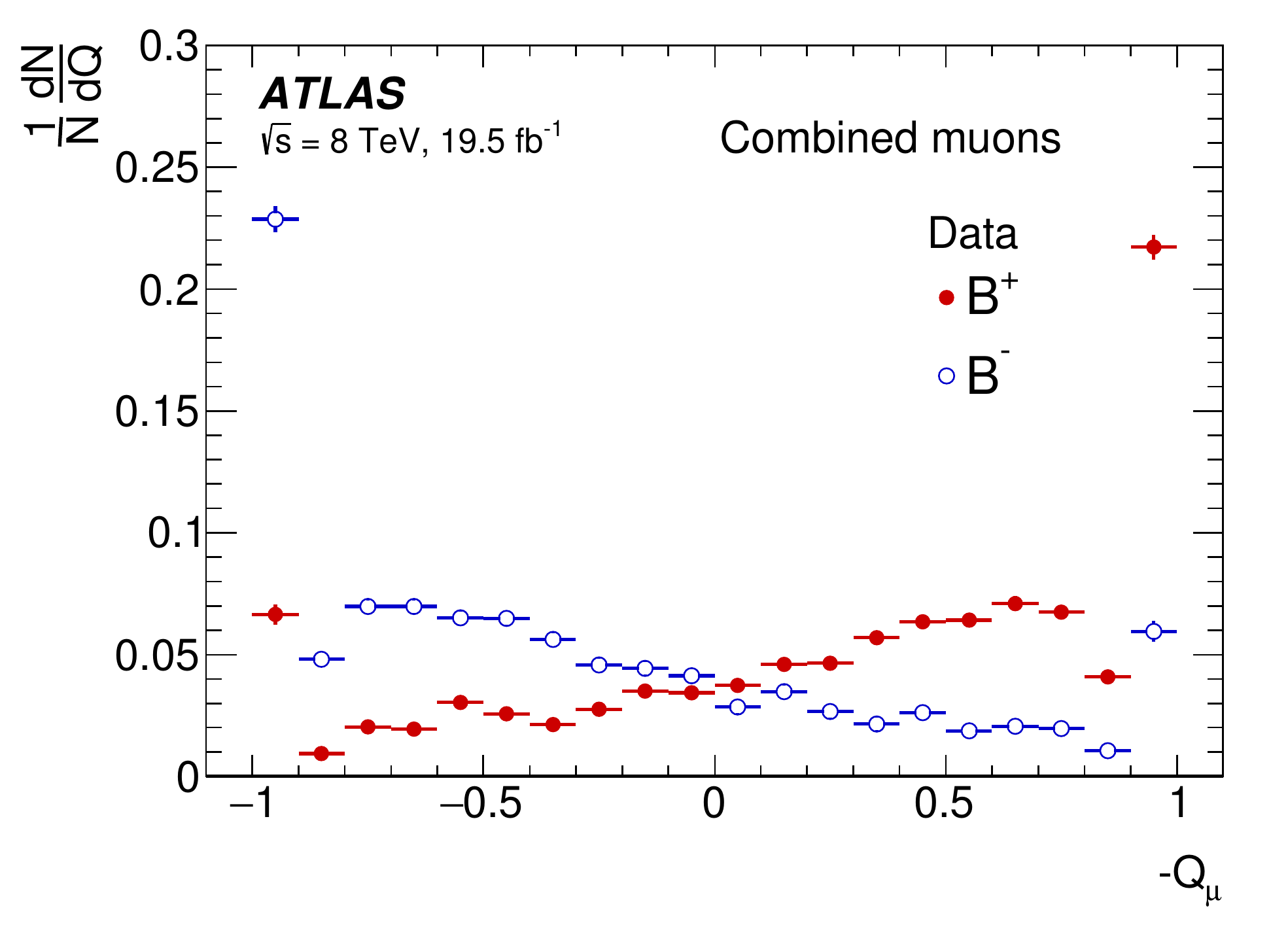}}
{
\ifthenelse {\boolean{Internal}}
{\includegraphics[width=0.45\textwidth]{tag_dist_mujet_tag_Internal.pdf}
\includegraphics[width=0.45\textwidth]{tag_dist_mujet_comb_Internal.pdf}}
{\includegraphics[width=0.45\textwidth]{tag_dist_mujet_tag_Preliminary.pdf}
\includegraphics[width=0.45\textwidth]{tag_dist_mujet_comb_Preliminary.pdf}}
}

\caption{The opposite-side muon cone charge distribution for $B^{\pm}$ signal candidates for  {\it segment-tagged} (left) and {\it combined} (right) muons. The $B^{\pm}$ charge is determined from the kaon charge.\label{fig:tagmudist}}
\end{center}
\end{figure}

A muon {\it cone charge} variable is constructed, defined as
\begin{equation*}
Q_{\mu}  =  \frac{ \sum^{N\;{\rm tracks}}_{i} q_{i} \cdot (p_{\mathrm{T} i })^{\kappa}} {  \sum^{N\;{\rm tracks}}_{i}(p_{\mathrm{T} i })^{\kappa}},
\end{equation*} 
where $q$ is the charge of the track, $\kappa=1.1$ and the sum is performed over the reconstructed ID tracks within a cone, $\Delta R = \sqrt{(\Delta\phi)^2+(\Delta\eta)^2}<0.5$, around the muon direction.  The reconstructed ID tracks must have $\pt>0.5$ \GeV\ and $|\eta|<2.5$. Tracks associated with the $B^{\pm}$ signal decay are excluded from the sum. In Figure~\ref{fig:tagmudist} the opposite-side muon cone charge distributions are shown for candidates from $B^{\pm}$ signal decays.

\begin{figure}[h]
\begin{center}
\ifthenelse {\boolean{Nothing}}
{\includegraphics[width=0.45\textwidth]{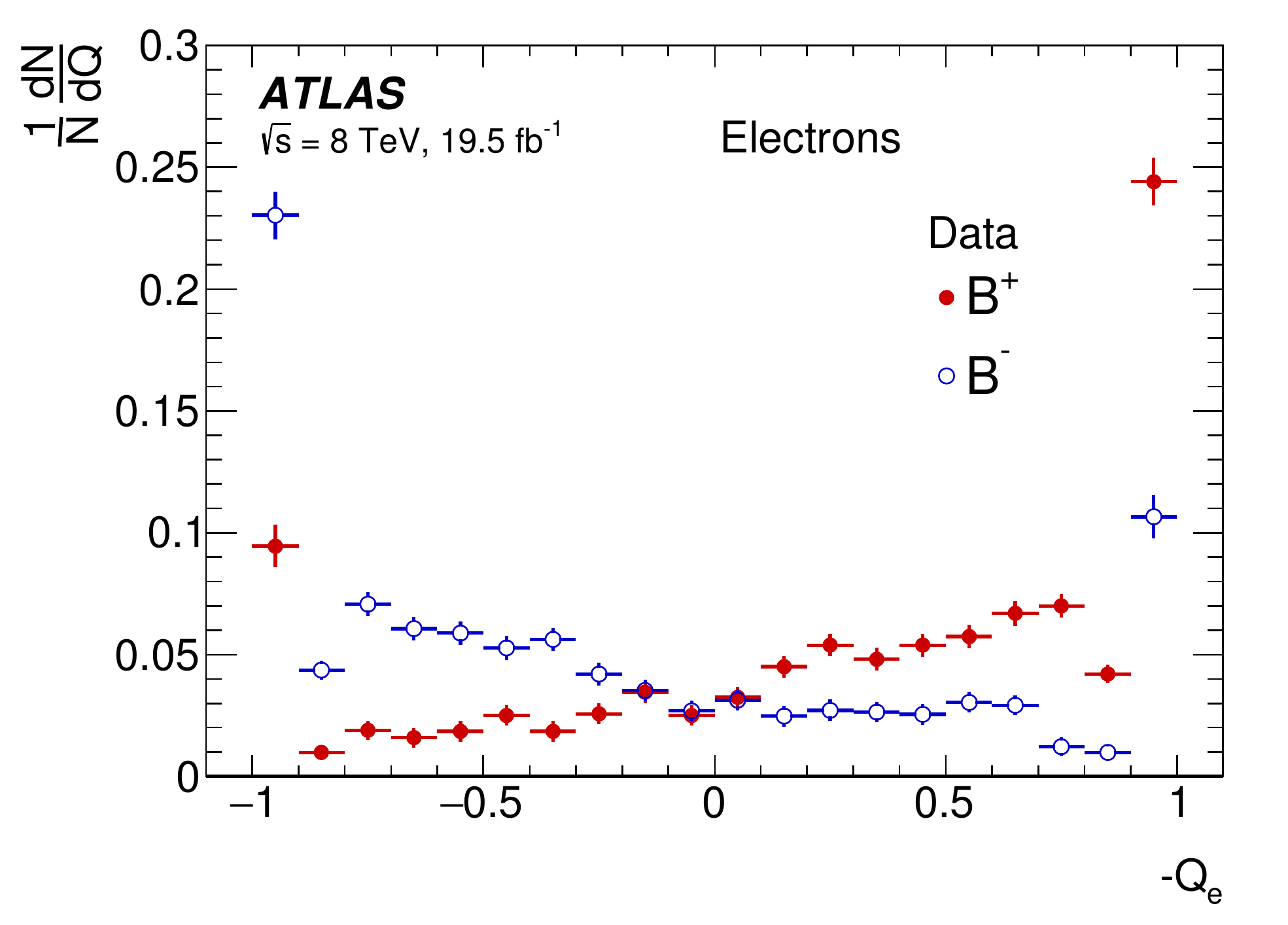}}
{
\ifthenelse {\boolean{Internal}}
{\includegraphics[width=0.45\textwidth]{tag_dist_eljet_Internal.pdf}}
{\includegraphics[width=0.45\textwidth]{tag_dist_eljet_Preliminary.pdf}}
}
\caption{The opposite-side electron cone charge distribution for $B^{\pm}$ signal candidates.\label{fig:tageldist}}
\end{center}
\end{figure}

For electron-based tagging, an electron is identified using information from the inner detector and calorimeter and is required to satisfy the tight electron quality criteria~\cite{electro1}. The inner detector track associated with the electron is required to have $\pT>0.5$ \GeV\ and $|\eta|<2.5$. It is required to pass within $|\Delta z|<5$~mm of the primary vertex to remove electrons from non-signal interactions. To exclude electrons associated with the signal-side of the decay,  electrons are rejected that have momenta within a cone of size $\Delta R = 0.4$ around the signal $B$ candidate direction in the laboratory frame and opening angle between the $B$ candidate and electron momenta,  $\zeta_{b}$,  of  $\cos(\zeta_{b})>0.98$. In the case of more than one electron passing the selection, the electron with the highest transverse momentum is chosen. As in the case of muon tagging, additional tracks within  a cone of size $\Delta R = 0.5$ are used to form the electron cone charge $Q_{e}$ with $\kappa=1.0$. If there are no additional tracks within the cone, the charge of the electron is used.
The resulting opposite-side electron cone charge distribution is shown in Figure~\ref{fig:tageldist} for $B^{+}$ and $B^{-}$ signal events.

In the absence of a muon or electron, {\it $b$-tagged} jets (i.e. jets that are the product of a $b$-quark) are identified using  a multivariate tagging algorithm~\cite{ATLAS-CONF-2014-046}, which is a combination of several $b$-tagging algorithms using an artificial neural network and outputs a {\it b-tag} weight classifier. Jets are selected that exceed a {\it b-tag} weight of 0.7.  This value is optimized to maximize the tagging power of the calibration sample. Jets are reconstructed from track information using the anti-${k_t}$ algorithm~\cite{Cacciari:2008gp} with a radius parameter $R = 0.8$. In the case of multiple jets, the jet with the highest value of the $b$-tag weight is used. 

\begin{figure}[h]
\begin{center}
\ifthenelse {\boolean{Nothing}}
{\includegraphics[width=0.45\textwidth]{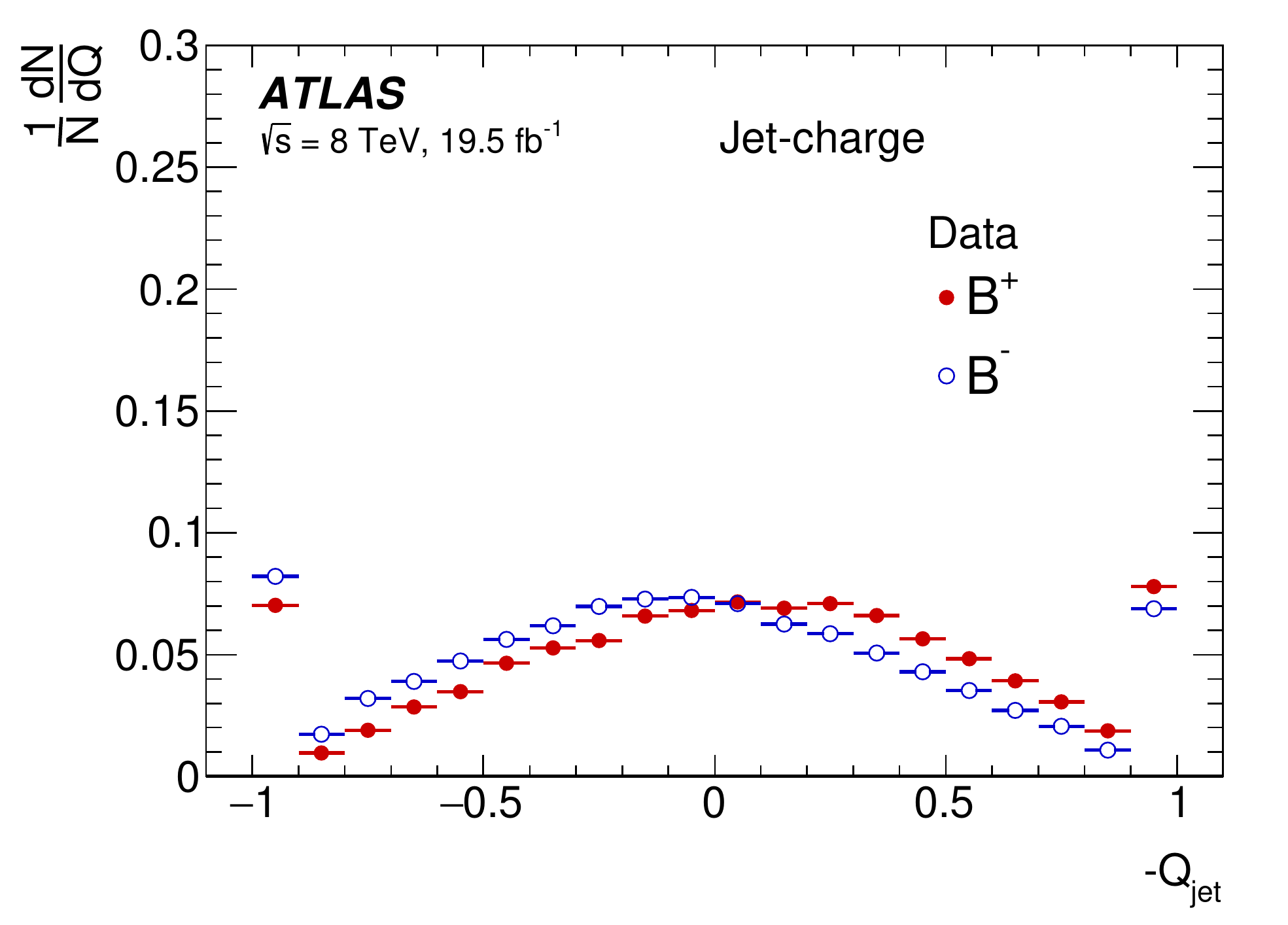}}
{
\ifthenelse {\boolean{Internal}}
{\includegraphics[width=0.45\textwidth]{tag_dist_jet_btag_Internal.pdf}}
{\includegraphics[width=0.45\textwidth]{tag_dist_jet_btag_Preliminary.pdf}}
}
\caption{Opposite-side jet-charge distribution for $B^{\pm}$ signal candidates.
\label{fig:jetchargedist}}
\end{center}
\end{figure}

The {\it jet charge} is defined as
\begin{equation*}
Q_{\rm jet}  =  \frac{ \sum^{N\;{\rm tracks}}_{i} q_{i} \cdot (p_{\mathrm{T} i })^{\kappa}} {  \sum^{N\;{\rm tracks}}_{i}(p_{\mathrm{T} i })^{\kappa}},
\end{equation*}
where $\kappa=1.1$ and the sum is over the tracks associated with the jet, excluding those tracks associated with a primary vertex other than that of the signal decay and tracks from the signal candidate.
Figure~\ref{fig:jetchargedist} shows the distribution of the opposite-side jet-charge for  $B^{\pm}$ signal candidates.

The efficiency, $\epsilon$, of an individual tagging method is defined as the ratio of the number of events tagged by that method to the total number of candidates.
 A probability $P(B|Q)$ ($P({\bar B}|Q)$) that a specific event has a signal decay containing a ${\bar b}$-quark ($b$-quark) given the value of the discriminating variable is constructed from the calibration samples for each of the $B^{+}$ and $B^{-}$ samples, which defines $P(Q|B^{+})$ and $P(Q|B^{-})$, respectively.  The probability to tag a signal event as containing a ${\bar b}$-quark is therefore $P(B|Q) = P(Q|B^{+}) / (P(Q|B^{+}) + P(Q|B^{-}))$, and correspondingly  $P({\bar B}|Q) = 1-P(B|Q)$.
 It is possible to define a quantity called the dilution ${\mathcal D}= P(B|Q) - P({\bar B}|Q) = 2 P({B}|Q) - 1$, which represents the strength of a particular flavour tagging method.  The tagging power of a particular tagging method is defined as $T=\epsilon{\mathcal D}^2 = \sum_{i} \epsilon_{i} \cdot (2 P_{i}({B}|Q_{i}) - 1)^{2}$, where the sum is over the bins of the probability distribution as a function of the charge variable. An effective dilution, $D=\sqrt{T/\epsilon}$, is calculated from the measured tagging power and efficiency.

The flavour tagging method applied to each \Bs \  candidate event is taken from the information contained in a given event. By definition there is no overlap between lepton-tagged and jet-charge-tagged events. The overlap between muon- and electron-tagged events, corresponding to 0.4\% of all tagged events, is negligibly small. In the case of doubly tagged events, the tagger with the highest tagging power is selected; however, the choice of hierarchy between muon- and electron-tagged events is shown to have negligible impact on the final fit results.
If it is not possible to provide a tagging response for the event, then a probability of 0.5 is assigned. A summary of the tagging performance is given in Table~\ref{tab:taggingresults}.

\begin{table}[t]
\begin{center}
\begin{tabular}{l|c|c|c}
\hline
Tagger & Efficiency $[\%]$ & Dilution $[\%]$ & Tagging Power $[\%]$ \\
\hline\hline
Combined $\mu$             & $ 4.12   \pm 0.02 $ & $ 47.4  \pm 0.2  $   & $ 0.92 \pm 0.02$ \\
Electron            & $ 1.19 \pm 0.01 $ & $ 49.2  \pm 0.3 $    & $ 0.29 \pm 0.01$ \\
Segment-tagged $\mu$  & $ 1.20 \pm 0.01 $ & $ 28.6  \pm 0.2 $    & $ 0.10 \pm 0.01$ \\
Jet-charge                       &  $ 13.15   \pm 0.03 $ & $ 11.85  \pm 0.03 $  & $ 0.19 \pm 0.01$ \\
\hline
Total &                            $ 19.66 \pm 0.04 $ & $ 27.56 \pm 0.06 $ & $ 1.49 \pm 0.02 $ \\
\hline
\end{tabular}
\end{center}
\caption{Summary of tagging performance for the different flavour tagging methods described in the text. Uncertainties shown are statistical only.  The efficiency and tagging power are each determined by summing over the individual bins of the charge distribution. The effective dilution is obtained from the measured efficiency and tagging power. For the efficiency, dilution, and tagging power, the corresponding uncertainty is determined by combining the appropriate uncertainties in the individual bins of each charge distribution.}
\label{tab:taggingresults}
\end{table}

\subsection{Using tag information  in the $B_s^0$ fit} \label{TagInFit}
The tag-probability for each \Bs\ candidate is determined from calibrations derived from a sample of $B^{\pm}\to J/\psi K^{\pm}$ candidates, as described in Section~\ref{sec:FTMethods}.  The distributions of tag-probabilities for the signal and background are different and  since  the background  cannot be factorized out, additional probability terms, $P_\textrm{s}(P(B|Q))$ and $P_\textrm{b}(P(B|Q))$ for signal and background, respectively, are included in the fit. The distributions of tag-probabilities for the $B_s^0$ candidates consist of continuous and discrete parts (events with a tag charge of $\pm 1$); these are treated separately as described below. 

To describe the continuous part, a fit is first performed to the sideband data,  i.e., $5.150$ \GeV\ $< m(B_s^0) <5.317$ \GeV\ or $5.417$ \GeV\ $< m(B_s^0) < 5.650$ \GeV, where $m(B_s^0)$ is the mass of the \Bs\ candidate.  Different functions are used for the different tagging methods. For the combined-muon tagging method, the function has the form of the sum of a fourth-order polynomial and two exponential functions.  A second-order polynomial and two exponential functions are applied for the electron tagging algorithm. A sum  of three Gaussian functions is used for the segment-tagged muons.  For the jet-charge tagging algorithm an eighth-order polynomial is used. In all four cases unbinned maximum-likelihood fits to data are used. In the next step, the same function as applied to the sidebands is  used to describe the distributions for events in the  signal region: the background  parameters are fixed to the values obtained from the fits to the sidebands while the signal parameters are free in this step.   The  ratio of  background to signal (obtained from a simultaneous mass--lifetime fit) is fixed as well.   The results of the fits projected onto histograms  of   \Bs\  tag-probability for the different tagging methods are shown in Figure~\ref{Fig:taggerfit}.

To account for possible deviations between data and the selected fit models a number of alternative fit functions are used to determine systematic uncertainties in the $B_s^0$ fit. These fit variations are described in Section \ref{sec:Systematics}.

\ifthenelse {\boolean{Nothing}}
{\begin{figure}[htb]
	\begin{center}
		\includegraphics[width=0.44\textwidth]{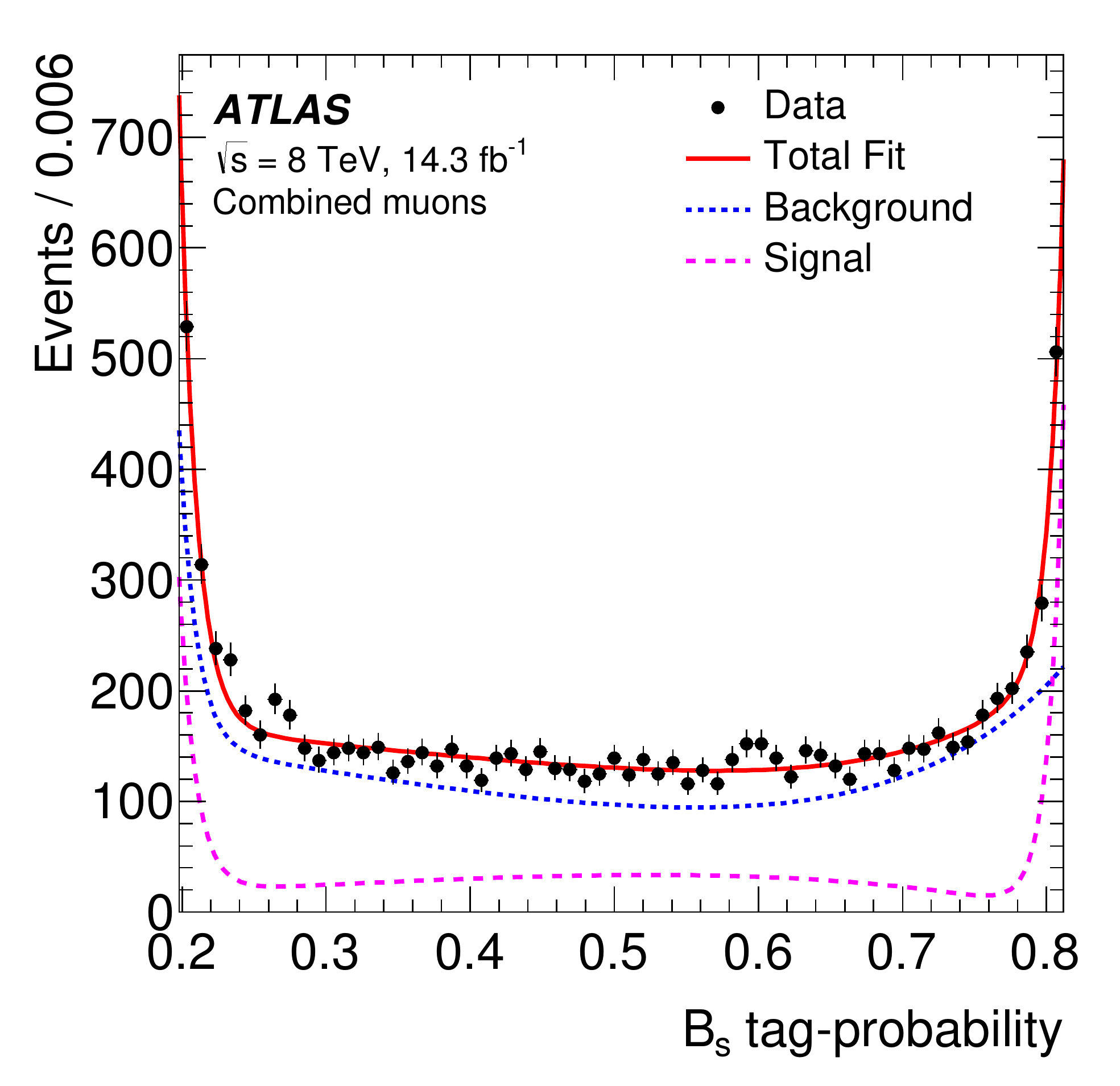}
		\includegraphics[width=0.44\textwidth]{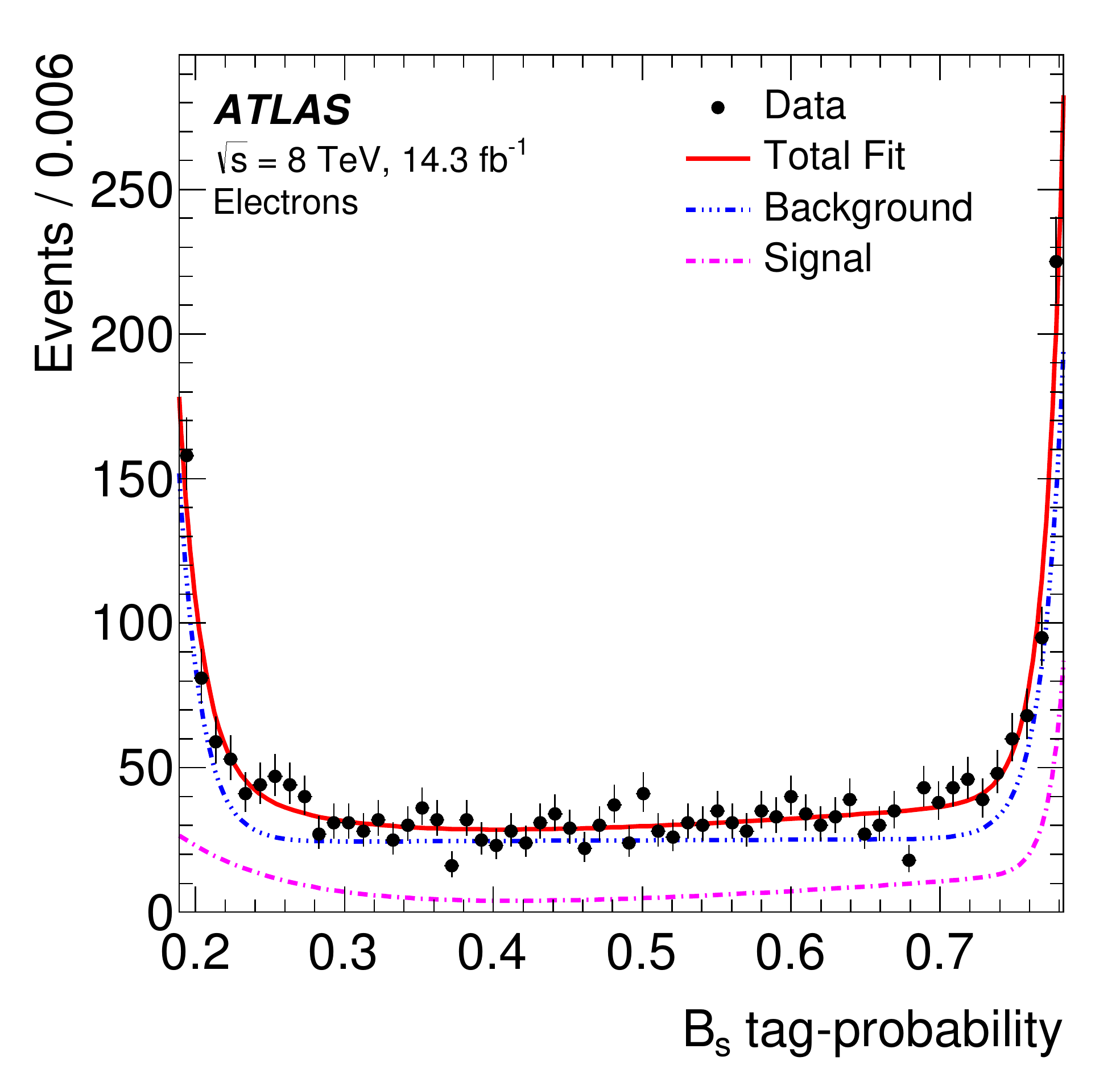}
		\includegraphics[width=0.44\textwidth]{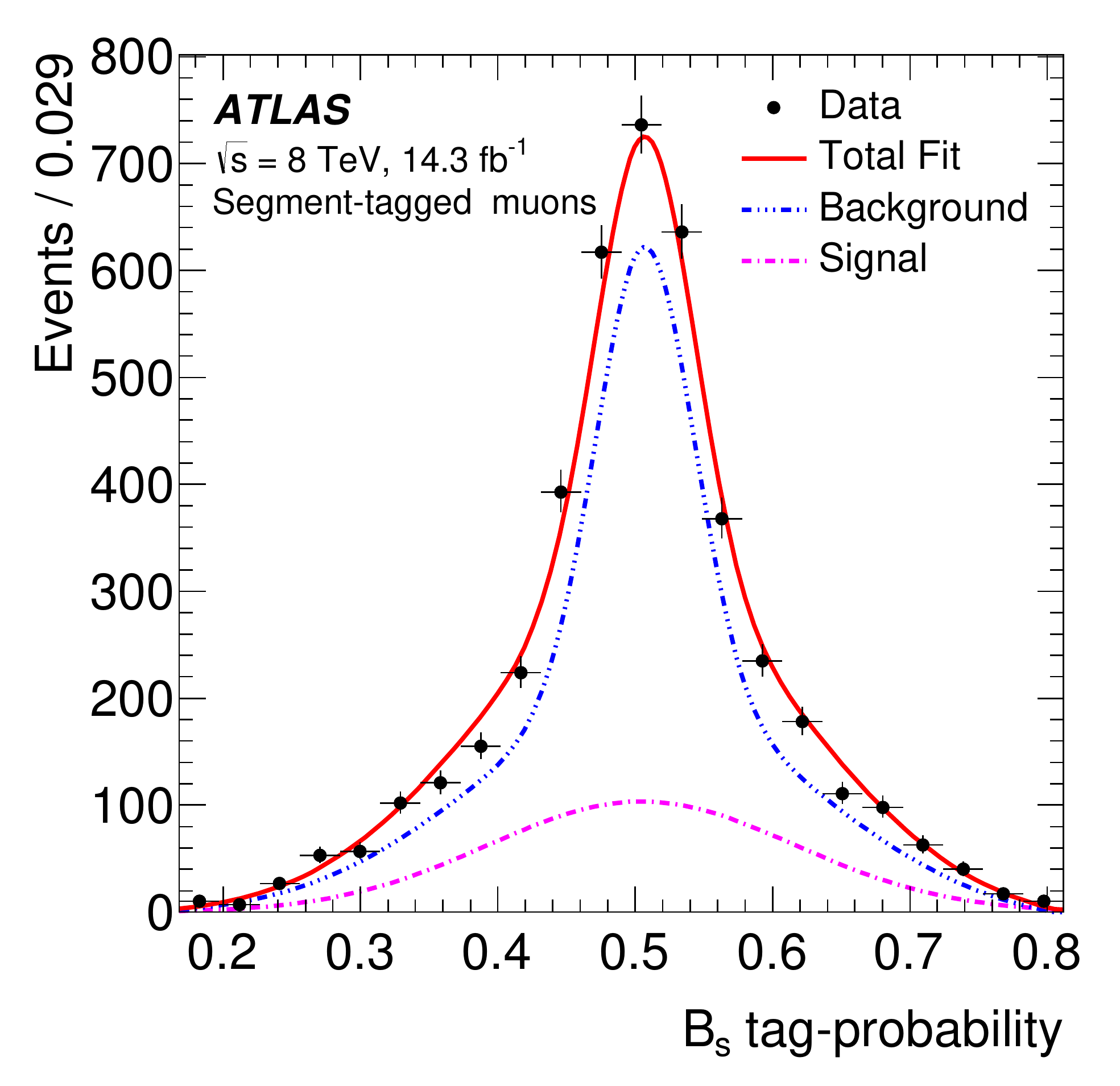}
		\includegraphics[width=0.44\textwidth]{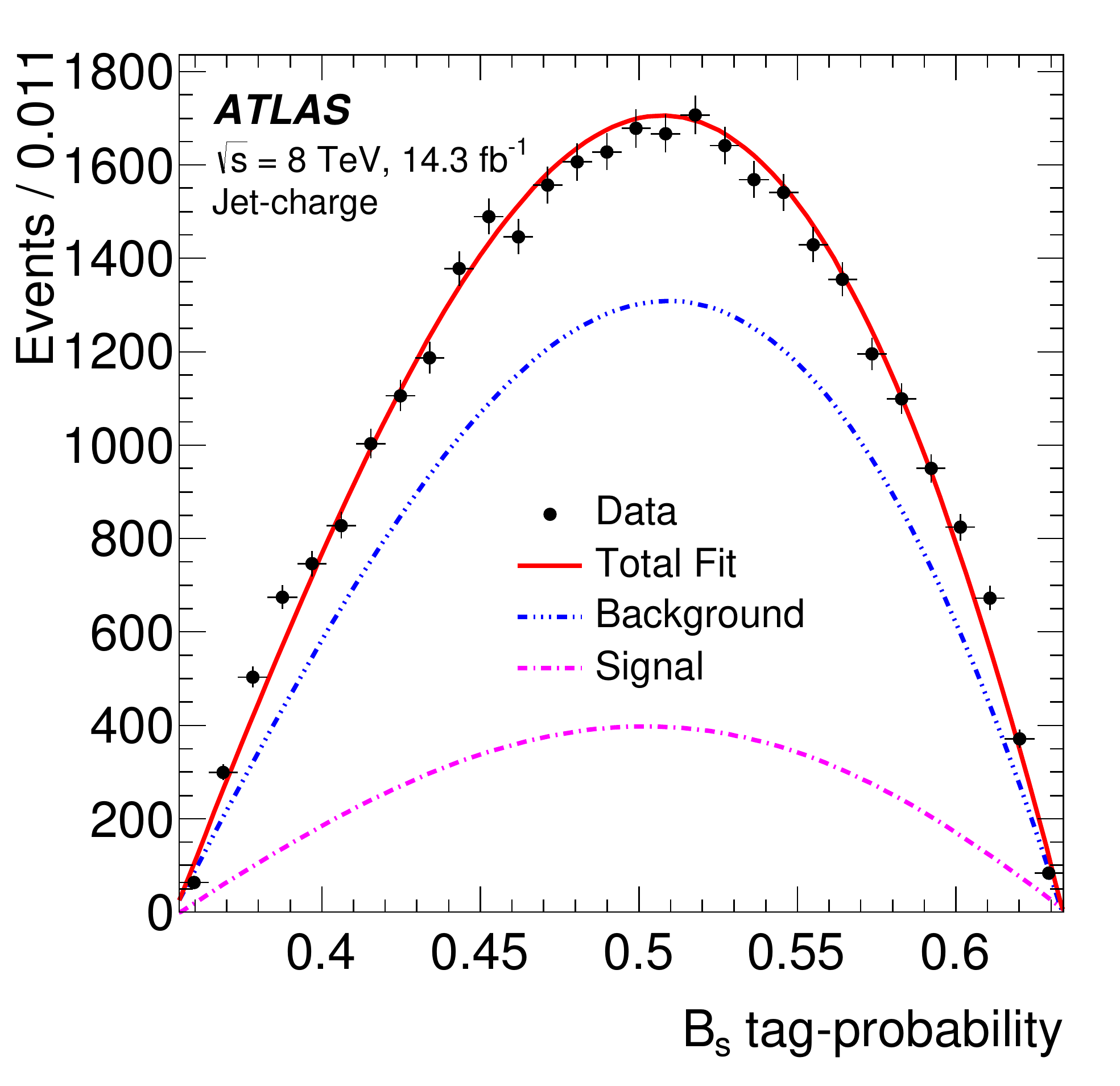}	
		\caption{The continuous part of tag-probability for tagging using combined-muons  (top-left), electrons (top-right), segment-tagged muons (bottom-left) and jet-charge  (bottom-right). Black dots are data, blue is a fit to the sidebands, purple to the signal and red is a sum of both fits.\label{Fig:taggerfit}}
	\end{center}
\end{figure}}
{
\ifthenelse {\boolean{Internal}}
{\begin{figure}[htb]
	\begin{center}
		\includegraphics[width=0.44\textwidth]{tag1fit_sig-green+bck-blue_Internal.pdf}
		\includegraphics[width=0.44\textwidth]{tag5fit_sig-green+bck-blue_Internal.pdf}
		\includegraphics[width=0.44\textwidth]{tag2fit_sig-green+bck-blue_Internal.pdf}
		\includegraphics[width=0.44\textwidth]{tag3fit_sig-green+bck-blue_Internal.pdf}	
		\caption{The continuous part of tag-probability for tagging using combined-muons  (top-left), electrons (top-right), segment-tagged muons (bottom-left) and jet-charge  (bottom-right). Black dots are data, blue is a fit to the sidebands, green to the signal and red is a sum of both fits.\label{Fig:taggerfit}}
	\end{center}
\end{figure}}
{\begin{figure}[htb]
	\begin{center}
		\includegraphics[width=0.44\textwidth]{tag1fit_sig-green+bck-blue_Preliminary.pdf}
		\includegraphics[width=0.44\textwidth]{tag5fit_sig-green+bck-blue_Preliminary.pdf}
		\includegraphics[width=0.44\textwidth]{tag2fit_sig-green+bck-blue_Preliminary.pdf}
		\includegraphics[width=0.44\textwidth]{tag3fit_sig-green+bck-blue_Preliminary.pdf}	
		\caption{The continuous part of tag-probability for tagging using combined-muons  (top-left), electrons (top-right), segment-tagged muons (bottom-left) and jet-charge  (bottom-right). Black dots are data, blue is a fit to the sidebands, green to the signal and red is a sum of both fits.\label{Fig:taggerfit}}
	\end{center}
\end{figure}}
}

The discrete components of the tag-probability distribution originate from cases where the tag is derived from a single track, giving a tag charge of exactly $+1$ or $-1$. 
The  fractions of events $f_{+1}$  and $f_{-1}$  with  charges $+1$ and  $-1$, respectively, 
are determined separately for signal and background using events from the same \Bs\ mass signal and sideband regions.
Positive and negative charges are equally probable for background candidates formed from a random combination of a 
$J/\psi$  and a pair of tracks,  but this is not the case for background candidates formed from  a partially reconstructed $b$-hadron. 
For signal and background contributions, similar fractions of events that are tagged with +1 or $-1$ tagging charge are observed for each of the tagging methods.
The remaining fraction of events, $1 - f_{+1} - f_{-1}$,   constitute the continuous part of the distributions. 
Table \ref{tab:TagPunziSpkes} summarizes the fractions $f_{+1}$  and $f_{-1}$ obtained for signal and 
background events and for the different tag methods.
\begin{table}[!h]
\begin{center}
\begin{tabular}{l|cc|cc}
  \hline
  Tag method & \multicolumn{2}{c|}{Signal} & \multicolumn{2}{c}{Background} \\
             & $f_{+1}$ & $f_{-1}$           & $f_{+1}$ & $f_{-1}$ \\
  \hline \hline

  Combined $\mu$       & $0.124 \pm 0.012$ & $0.127 \pm 0.012$ & $0.093 \pm 0.003$ & $0.095 \pm 0.003$ \\
  Electron             & $0.105 \pm 0.020$ & $0.139 \pm 0.021$ & $0.110 \pm 0.007$ & $0.110 \pm 0.007$ \\
  Segment-tagged $\mu$ & $0.147 \pm 0.024$ & $0.118 \pm 0.023$ & $0.083 \pm 0.004$ & $0.084 \pm 0.004$ \\
  Jet-charge           & $0.071 \pm 0.005$ & $0.069 \pm 0.005$ & $0.068 \pm 0.002$ & $0.069 \pm 0.002$ \\

  \hline

\end{tabular}
\end{center}
\caption{Table summarizing the fraction of events $f_{+1}$  and $f_{-1}$  with tag charges of $+1$ and $-1$, respectively for signal and background events and for the different  tag methods. Only statistical errors are quoted. }
\label{tab:TagPunziSpkes}

\end{table}

\begin{table}[!h]
\begin{center}
\begin{tabular}{l|c|c}
 \hline
 Tag method & Signal      & Background \\
 \hline\hline
 Combined $\mu$    & $0.047 \pm 0.003  $ & $0.038  \pm 0.001 $ \\
  Electron                 & $0.012 \pm 0.001  $ & $0.008  \pm 0.001 $ \\
Segment-tagged $\mu$ & $0.013 \pm 0.001  $ & $0.015  \pm 0.001 $ \\
Jet-charge                & $0.135  \pm 0.003 $ & $0.100  \pm 0.001  $ \\
Untagged                & $0.793  \pm 0.002 $ & $0.839  \pm 0.002  $ \\
 \hline\hline
\end{tabular}
\end{center}
\caption{Table summarizing the relative fractions of signal and background events tagged 
using the different tag methods. The fractions include both the continuous and discrete contributions.
Only statistical errors are quoted. \label{tab:TagPunziRelat}}
\end{table}
To estimate the fractions of signal and background events which have tagging, a similar sideband--subtraction method is used to determine the relative fraction of signal and background events tagged using the different methods.
These fractions are also included in the maximum-likelihood fit, described in Section \ref{sec:PDF}. 
The results are summarized in Table \ref{tab:TagPunziRelat}.

%% file: Section5.tex
\section{Maximum likelihood fit \label{sec:PDF}}
An unbinned maximum-likelihood fit is performed on the selected events to extract the parameter values of the \Bsto\ decay.  The fit uses information about the reconstructed mass $m$, the measured proper decay time $t$, the measured proper decay time uncertainty $\sigma_t$,  the tagging probability, and the transversity angles $\Omega$ of each \Bst\ decay candidate.  The measured proper decay time uncertainty $\sigma_t$ is calculated from the covariance matrix associated with the vertex fit of each candidate event.  The transversity angles $\Omega = (\theta_T,\psi_T,\phi_T)$ are defined in Section \ref{sec:sigPDF}.
The likelihood is independent of the $K^+K^-$ mass distribution. 
The likelihood function is defined as a combination of the signal and background probability density functions as follows:
\begin{alignat}{3} \label{eq:likelihood}
\ln~{\cal L}  =  \sum_{i=1}^N \{   w_i \cdot  \mathrm{ln} (  f_{\textrm{s}} \cdot   {\cal F}_{\textrm{s}}(  m_{i}, t_{i}, \sigma_{t_{i}}, \Omega_i , P(B|Q), {\it p_{\mathrm{T}_{i}}} )
 +  f_{\textrm{s}} \cdot f_{B^0} \cdot {\cal F}_{B^0}  (  m_{i}, t_{i}, \sigma_{t_{i}}, \Omega_i , P(B|Q), {\it p_{\mathrm{T}_{i}}} ) \nonumber \\
 +  f_{\textrm{s}} \cdot f_{\Lambda_b} \cdot {\cal F}_{\Lambda_b}  (  m_{i}, t_{i}, \sigma_{t_{i}}, \Omega_i , P(B|Q), {\it p_{\mathrm{T}_{i}}} ) \nonumber \\
 \vphantom{\sum_{i=1}^N} + ( 1 -  f_{\textrm{s}} \cdot (1 + f_{B^0} + f_{\Lambda_b})    )  {\cal F}_{\textrm{bkg}} (  m_{i}, t_{i}, \sigma_{t_{i}}, \Omega_i, P(B|Q), {\it p_{\mathrm{T}_{i}}} )  )  \},   \nonumber \\
\end{alignat}
where $N$ is the number of selected candidates,  $w_i$ is a weighting factor to account for the trigger efficiency (described in Section~\ref{sec:Teff}), and $f_\textrm{s}$ is the fraction of signal candidates. The background fractions $f_{B^0}$ and $f_{\Lambda_b}$ are the fractions of $B^0$ mesons and $\Lambda_b$ baryons mis-identified as \Bs\ candidates calculated relative to the number of signal events; these parameters are fixed to their MC values and varied as part of the systematic uncertainties. The  mass $m_{i}$, the proper decay time $t_{i}$ and the decay angles $\Omega_i $ are the values measured from the data for each event $i$.  
${ \cal F}_{\textrm{s} }$, ${\cal F}_{B^0}$, ${\cal F}_{\Lambda_b}$ and  ${\cal F}_{\textrm{bkg} }$  are the probability density functions (PDF) modelling the signal, $B^0$  background, $\Lambda_b$ background, and the other background distributions, respectively.  A detailed description of the signal PDF terms in Equation~(\ref{eq:likelihood}) is given in Section \ref{sec:sigPDF}.  The three background functions are described in Section \ref{sec:bgPDF}.

\subsection{Signal PDF\label{sec:sigPDF}} 
The PDF used to describe  the signal events, ${\cal F}_{\textrm{s}}$, has the following composition:
\begin{eqnarray}\label{eq:likelihoodSig}
  {\cal F}_{\textrm{s}}(  {\it m_{i}, t_{i},} \sigma_{t_{i}}, \Omega_i, P(B|Q), {\it p_{\mathrm{T}_{i}}} ) & = & P_\textrm{s}(m_{i}) \cdot P_\textrm{s}(\Omega_i, t_{i}, P(B|Q), \sigma_{t_{i}}) \nonumber \\
 & & \cdot P_\textrm{s}(\sigma_{t_{i}}) \cdot P_\textrm{s}(P(B|Q)) \cdot A(\Omega_i,{\it p_{\mathrm{T}_{i}}})  \cdot P_\textrm{s}({\it p_{\mathrm{T}_{i}}}).
\end{eqnarray}
The mass function $P_\textrm{s}(m_{i})$ is modelled by a sum of three Gaussian distributions. 
The probability terms $P_\textrm{s}(\sigma_{t_{i}})$ and $P_\textrm{s}({\it p_{\mathrm{T}i}})$ are described by gamma functions and are unchanged from the analysis described in Ref.~\cite{ATLASBsJpsiphi}. The tagging probability term for signal $P_\textrm{s}(P(B|Q))$ is described in Section \ref{TagInFit}.  

The term $P_\textrm{s}(\Omega_i, t_{i}, P(B|Q), \sigma_{t_{i}}) $ is a joint PDF for the decay time $t$ and the transversity angles $\Omega$ for the \Bsto\ decay.  Ignoring detector effects, the distribution for the time $t$ and the angles $\Omega$  is given by the differential decay rate \cite{CDFAngles}:

\begin{equation*}\label{eq:ang2}
\frac{\rm{d}^4 \Gamma}{\rm{d}t\ \rm{d}\Omega}= \sum_{k=1}^{10} {\cal O}^{(k)}(t) g^{(k)}(\theta_T,\psi_T,\phi_T), 
\end{equation*}
where  ${\cal O}^{(k)}(t)$ are the time-dependent functions corresponding to the contributions of the four different amplitudes ($A_0$, $A_{||}$, $A_\perp$, and $A_S$) and their interference terms, and $g^{(k)}(\theta_T,\psi_T,\phi_T)$ are the angular functions.   Table \ref{tab:SigPDF} shows these time-dependent functions and the angular functions of the transversity angles.  The formulae for the time-dependent functions have the same structure for \Bs\ and \aBs\ but with a sign reversal in the terms containing $\Delta m_s$.  
In Table \ref{tab:SigPDF}, the parameter $A_\perp(t)$ is the time-dependent amplitude for the $CP$-odd final-state configuration while $A_0(t)$ and $A_\parallel(t)$ correspond to $CP$-even final-state configurations.  The amplitude $A_S(t)$ gives the contribution from the $CP$-odd non-resonant $B_s^0\rightarrow J/\psi K^+K^-$ $S$-wave state (which includes the $f_0$).  The corresponding functions are given in the last four lines of Table \ref{tab:SigPDF}  ($k=7$--$10$). The amplitudes are parameterized by $|A_i| e^{ i \delta_i }$, where $i = \{ 0, ||, \perp, S \}$, with $\delta_0 = 0$ and are normalized such that $|A_0(0)|^2 + |A_\perp(0)|^2 +  |A_{\|}(0)|^2=1$. $|A_\perp(0)|$ is determined according to this condition, while the remaining three amplitudes are parameters of the fit.  The formalism used throughout this analysis assumes no direct CP violation.

\begin{figure}
  \centering
  \ifthenelse {\boolean{Nothing}}
	      {\includegraphics[width=0.45\textwidth]{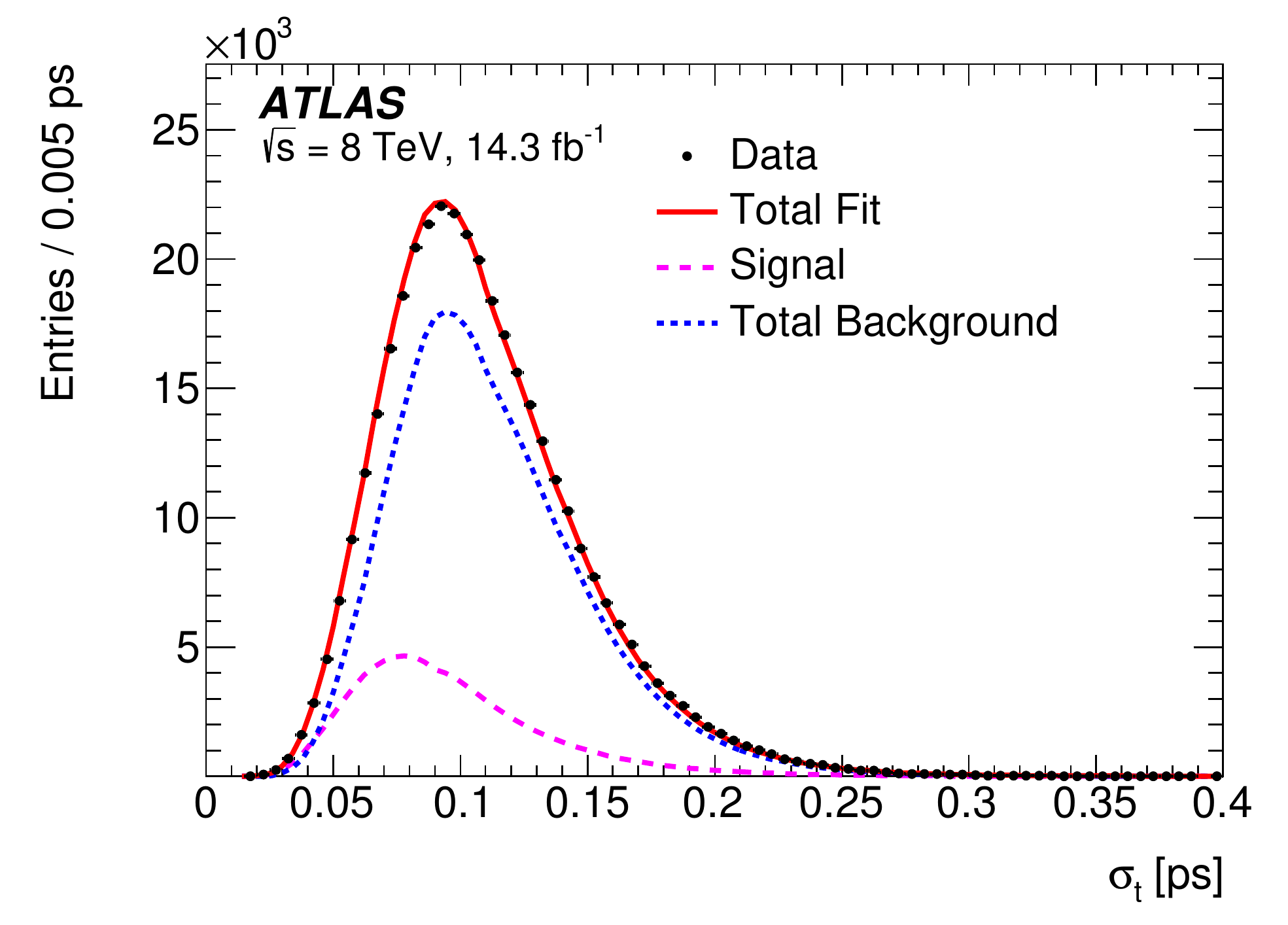}}	
              {
                \ifthenelse {\boolean{Internal}}
	                    {\includegraphics[width=0.45\textwidth]{timeErr_Internal.eps}}
	                    {\includegraphics[width=0.45\textwidth]{timeErr_Preliminary.eps}}
              }	
	      \caption{The proper decay time uncertainty distribution for data (black), and the fits to the background (blue) and the signal (purple) contributions. The total fit is shown as a red curve.}
	      \label{timeErr}
\end{figure}

The angles ($\theta_T,\psi_T,\phi_T$), are defined in the rest frames of the final-state particles.  The $x$-axis is determined by the direction of the $\phi$ meson in the $J/\psi$ rest frame, and the $K^{+}K^{-}$  system defines the $x$--$y$ plane, where $p_{y}(K^{+})>0$. The three angles are defined as:
\begin{itemize}
\item$\theta_T$, the angle between $\vec{p}(\mu^{+})$ and the normal to the $x$--$y$ plane, in the $J/\psi$ meson rest frame,
\item $\phi_T$, the angle between the $x$-axis and $\vec{p}_{xy}(\mu^{+})$, the projection of the $\mu^+$ momentum in the $x$--$y$ plane, in the $J/\psi$ meson rest frame,
\item $\psi_T$, the angle between $\vec{p}(K^{+})$ and $-\vec{p}(J/\psi)$ in the $\phi$ meson rest frame.
\end{itemize}
The  PDF term $P_\textrm{s}(\Omega_i, t_{i}, P(B|Q), \sigma_{t_{i}})$ takes into account the lifetime resolution, so each time element in Table \ref{tab:SigPDF} is smeared with a Gaussian function. This smearing is performed numerically on an event-by-event basis where the width of the Gaussian function is the proper decay time uncertainty, measured for each event, multiplied by a scale factor to account for any mis-measurements. The proper decay time uncertainty distribution for data,  including  the fits to the background and the signal contributions is shown in Figure~\ref{timeErr}. The average value of this uncertainty for signal events is 97 fs.

The angular acceptance of the detector and kinematic cuts on the angular distributions are included in the likelihood function through $A(\Omega_i, {\it p_{Ti}})$. This is calculated using a 4D binned acceptance method, applying an event-by-event efficiency according to the transversity angles ($\theta_T,\psi_T,\phi_T$) and the \pT\ of the candidate. The \pT\  binning is necessary, because the angular acceptance is influenced by the \pT\  of the \Bs\ candidate.  The acceptance is calculated from the \Bst\ MC events.
Taking the small discrepancies between data and MC events into account have negligible effect on the fit results. In the likelihood function, the acceptance is treated as an angular acceptance PDF, which is multiplied with the time- and angle-dependent PDF describing the \Bsto\ decays.  As both the acceptance and time- and angle-dependent decay PDFs depend on the transversity angles they must be normalized together. This normalization is done numerically during the likelihood fit. 
The PDF is normalized over the entire \Bs \ mass range \cutMassBsRange.

\begin{sidewaystable}[hp]
\begin{center}
\begin{tabular}{|c| l | l |}
\hline
$k$ & ${\cal O}^{(k)}(t)$ & $g^{(k)}(\theta_T,\psi_T,\phi_T)$ \\
\hline \hline
1 & $\frac{1}{2} |A_0(0)|^2 \left[\left(1+\cos\phis\right) e^{-\Gamma_{\rm L}^{(s)} t} + \left(1-\cos\phis\right) e^{-\Gamma_{\rm H}^{(s)} t} \pm 2 e^{-\Gamma_st} \sin (\Delta m_s t)  \sin \phi_s \right]$ & $2 \cos^2\psi_T(1 - \sin^2\theta_T\cos^2\phi_T)$ \\
2 & $\frac{1}{2}|A_{\|}(0)|^2 \left[\left(1+\cos\phis\right) e^{-\Gamma_{\rm L}^{(s)} t} + \left(1-\cos\phis\right) e^{-\Gamma_{\rm H}^{(s)} t} \pm 2 e^{-\Gamma_st} \sin (\Delta m_s t)  \sin \phi_s \right]$ & $\sin^2\psi_T(1 - \sin^2\theta_T\sin^2\phi_T)$ \\
3 & $\frac{1}{2} |A_{\perp}(0)|^2 \left[\left(1- \cos\phis\right)e^{-\Gamma_{\rm L}^{(s)} t} + \left(1+\cos\phis\right)e^{-\Gamma_{\rm H}^{(s)} t} \mp 2 e^{-\Gamma_st} \sin (\Delta m_s t)  \sin \phi_s \right]$ & $\sin^2\psi_T\sin^2\theta_T$ \\

4 & $\frac{1}{2}|A_0(0)||A_{\|}(0)|\cos\delta_{||} $ & $  \frac{1}{\sqrt{2}}\sin2\psi_T\sin^2\theta_T\sin2\phi_T $ \\
  & \hfill $\left[\left(1+\cos\phis\right)e^{-\Gamma_{\rm L}^{(s)} t} + \left(1-\cos\phis\right)e^{-\Gamma_{\rm H}^{(s)} t} \pm 2 e^{-\Gamma_st} \sin (\Delta m_s t)  \sin \phi_s\right]$&  \\
  
5 & $|A_{\|}(0)||A_{\perp}(0)| [  \frac{1}{2}(e^{-\Gamma_{\rm L}^{(s)} t}-e^{-\Gamma_{\rm H}^{(s)} t})\cos(\delta_{\perp} - \delta_{||})\sin\phis$ &  $- \sin^2\psi_T\sin2\theta_T\sin\phi_T$ \\

 & \hfill $\pm e^{-\Gamma_st}(\sin (\delta_\perp -\delta_\|) \cos(\Delta m_s t) - \cos (\delta_\perp - \delta_\|)\cos \phi_s \sin (\Delta m_st) ) ]$ & \\
 6 & $|A_{0}(0)||A_{\perp}(0)| [  \frac{1}{2}(e^{-\Gamma_{\rm L}^{(s)} t}-e^{-\Gamma_{\rm H}^{(s)} t})\cos \delta_{\perp} \sin\phis$ & $\frac{1}{\sqrt{2}} \sin2\psi_T\sin2\theta_T\cos\phi_T$ \\
 & \hfill $\pm e^{-\Gamma_st}(\sin \delta_\perp \cos(\Delta m_s t) - \cos \delta_\perp \cos \phi_s \sin (\Delta m_st) ) ]$ & \\
7 &$\frac{1}{2} |A_{S}(0)|^2 \left[\left(1- \cos\phis\right)e^{-\Gamma_{\rm L}^{(s)} t} + \left(1+\cos\phis\right)e^{-\Gamma_{\rm H}^{(s)} t} \mp 2 e^{-\Gamma_st} \sin (\Delta m_s t)  \sin \phi_s \right]$ &$ \frac{2}{3}\left(1-\sin^2\theta_T\cos^2\phi_T\right) $ \\
 8 & $|A_{S}(0)||A_{\parallel}(0)| [  \frac{1}{2}(e^{-\Gamma_{\rm L}^{(s)} t}-e^{-\Gamma_{\rm H}^{(s)} t})\sin (\delta_\| - \delta_S) \sin\phis$ & $\frac{1}{3} \sqrt{6} \sin\psi_T\sin^2\theta_T\sin 2\phi_T$ \\
 & \hfill $\pm e^{-\Gamma_st}(\cos ( \delta_\| - \delta_S ) \cos(\Delta m_s t) - \sin ( \delta_\| - \delta_S ) \cos \phi_s \sin (\Delta m_st) ) ]$ & \\
9 &$\frac{1}{2}|A_{S}(0)| |A_{\perp}(0)|\sin(\delta_{\perp}-\delta_{S}) $ & $\frac{1}{3} \sqrt{6} \sin\psi_T\sin 2\theta_T\cos\phi_T$ \\
  &\hfill $\left[\left(1- \cos\phis\right)e^{-\Gamma_{\rm L}^{(s)} t} + \left(1+\cos\phis\right)e^{-\Gamma_{\rm H}^{(s)} t} \mp 2 e^{-\Gamma_st} \sin (\Delta m_s t)  \sin \phi_s \right]$ &  \\
 10 & $|A_{0}(0)||A_{S}(0)| [  \frac{1}{2}(e^{-\Gamma_{\rm H}^{(s)} t}-e^{-\Gamma_{\rm L}^{(s)} t})\sin  \delta_S \sin\phis$ & $\frac{4}{3}\sqrt{3}\cos\psi_T\left(1-\sin^2\theta_T\cos^2\phi_T\right)$ \\
 & \hfill $\pm e^{-\Gamma_st}(\cos \delta_S \cos(\Delta m_s t) + \sin \delta_S \cos \phi_s \sin (\Delta m_st) ) ]$ & \\
\hline
\end{tabular}
\end{center}
\caption{Table showing the ten time-dependent functions, ${\cal O}^{(k)}(t)$ and the functions of the transversity angles $g^{(k)}(\theta_T,\psi_T,\phi_T)$. The amplitudes $|A_0(0)|^2$ and $|A_\parallel(0)|^2$ are for the $CP$-even components of the \Bst\ decay,  $|A_{\perp}(0)|^2$ is the $CP$-odd amplitude; they have corresponding strong phases $\delta_{0}$, $\delta_{\parallel}$ and $\delta_{\perp}$. By convention $\delta_0$ is set to be zero.  The $S$-wave amplitude $|A_S(0)|^2$ gives the fraction of \BstoKK\ and has a related strong phase $\delta_S$. The $\pm$ and $\mp$ terms denote two cases: the upper sign describes the decay of a meson that was initially a \Bs\ meson, while the lower sign describes the decays of a meson that was initially \aBs. \label{tab:SigPDF}}
\end{sidewaystable}%

\clearpage

\subsection{Background PDF\label{sec:bgPDF}}
The background PDF has the following composition:
\begin{eqnarray} \label{eq:likelihoodBkg}
  {\cal F}_{\textrm{bkg}}(  {\it m_{i}, t_{i}}, \sigma_{t_{i}}, \Omega_i, P(B|Q), {\it p_{\mathrm{T}_{i}}} ) & =  & P_\textrm{b}(m_{i})\cdot  P_\textrm{b}(t_{i}|\sigma_{t_{i}}) \cdot P_\textrm{b}(P(B|Q))  \nonumber \\ 
                                                            & & \cdot P_\textrm{b}(\Omega_i)\cdot P_\textrm{b}(\sigma_{t_{i}})  \cdot P_\textrm{b}({\it p_{\mathrm{T}i}}).
\end{eqnarray}
The proper decay time function $P_\textrm{b}(t_{i}|\sigma_{t_{i}})$ is parameterized as a prompt peak modelled by a Gaussian distribution, two positive exponential functions and a negative exponential function. These functions are smeared with the same resolution function as the signal decay time-dependence. The prompt peak models the combinatorial background events, which are expected to have reconstructed lifetimes distributed around zero. The two positive exponential functions represent a  fraction of longer-lived backgrounds with non-prompt $J/\psi$, combined with hadrons from the primary vertex or  from a  $ B/D$ meson in the same event.  The negative exponential function takes into account events with poor vertex resolution.  
The probability terms $P_\textrm{b}(\sigma_{t_{i}})$ and $P_\textrm{b}({\it p_{\mathrm{T}i}})$ are described by gamma functions. They are unchanged from the  analysis described in  Ref.~\cite{ATLASBsJpsiphi}  and explained in detail there.  The tagging probability term for background $P_\textrm{b}(P(B|Q))$ is described in Section \ref{TagInFit}. 

The shape of the background angular distribution, $P_\textrm{b}(\Omega_i)$ arises primarily from detector and kinematic acceptance effects.  These are  described by Legendre polynomial functions: 
\begin{eqnarray} \label{eq:likelihoodBkgAng}
  Y^m_l (\theta_T) & = &  \sqrt{(2l+1)/(4\pi)}\sqrt{(l-m)!/(l+m)!}P^{|m|}_l(\cos\theta_T)  \nonumber  \\
   P_k(x)  & =  & \frac{1}{2^k k!} \frac{\rm{d}^k}{\rm{d}x^k}(x^2-1)^k   \\
 \mathscr{P}_{\mathrm{b}}(\theta_T,\psi_T,\phi_T)  & = &  \sum_{k=0}^{6}\sum_{l=0}^{6}\sum_{m=-l}^{l}
\begin{cases}
      a_{k,l,m}\sqrt{2} Y^m_l(\theta_T) \cos(m\phi_T) P_k(\cos\psi_T)  & \textrm{where } m > 0 \nonumber \\
			a_{k,l,m}\sqrt{2} Y^{-m}_l(\theta_T) \sin(m\phi_T) P_k(\cos\psi_T)  & \textrm{where } m < 0 \nonumber\\
			a_{k,l,m}\sqrt{2} Y^{0}_l(\theta_T)  P_k(\cos\psi_T)  & \textrm{where } m = 0 \nonumber \\
\end{cases}
\end{eqnarray}
where the coefficients $a_{k,l,m}$  are adjusted to give the best fit to the angular distributions for events in the \Bs\ mass sidebands.  These parameters are then fixed in the main fit. 
The \Bs \ mass interval used for the background fit is between 5.150  and  5.650 \GeV\ excluding the signal mass region $|(m(B_s^0) -5.366$ \GeV $|< 0.110$ \GeV. 
 The background mass model, $P_\textrm{b}({\it m_{i}})$ is an exponential function with a constant term added.
 
 Contamination from $B_d \ra J/\psi K^{0*}$ and $\Lambda_b \ra J/\psi p K^{-}$ events mis-reconstructed as \Bst\ are accounted for in the fit through the ${\cal F}_{B^0}$ and ${\cal F}_{\Lambda_b}$ terms in the PDF function described in Equation~(\ref{eq:likelihood}). The fraction of these contributions, $f_{B^0} = \Bdfrac$ and $f_{\Lambda_b} = (1.8 \pm 0.6)\%$, are evaluated from  MC simulation using production and branching fractions from \ifthenelse {\boolean{BdFraction3.3}}{Refs.~\cite{PDG,LHCb-CONF-2013-011,BABAR,LHCbLambdabProduction,1674-1137-40-1-011001,PhysRevLett.115.072001}}{Refs.~\cite{Dorigo:2013xf,LHCb-CONF-2013-011,BABAR,PhysRevLett.115.072001,PhysRevLett.111.102003}}. MC simulated events are also used to determine the shape of the mass and transversity angle distributions. The 3D angular distributions of $B_d^0\rightarrow J/\psi K^{*0}$ and of the conjugate decay are modelled using input from  Ref.~\cite{LHCbK0Star}, while angular distributions for $\Lambda_b \ra J/\psi p K^{-}$ and the conjugate decay are modelled as flat. These distributions are sculpted for detector acceptance effects and then described by Legendre polynomial functions, Equation~(\ref{eq:likelihoodBkgAng}), as in the case of the background described by Equation~(\ref{eq:likelihoodBkg}). These shapes are fixed in the fit. The $B_d$ and $\Lambda_b$ lifetimes are accounted for in the fit by adding additional exponential terms, scaled by the  ratio of $B_d$/$B_s^0$ or $\Lambda_b$/$B_s^0$ masses as appropriate, where the lifetimes and masses are taken from Ref.~\cite{PDG}.  Systematic uncertainties due to the background from $B_d \ra J/\psi K^{0*}$ and $\Lambda_b \ra J/\psi p K^{-}$ decays are described in Section \ref{sec:Systematics}. The contribution of $B_d \ra J/\psi K\pi$ events as well as their interference with $B_d \ra J/\psi K^{0*}$ events is not included in the fit and is instead assigned as a systematic uncertainty.

To account for possible deviations between data and the selected fit models a number of alternative fit functions and mass selection criteria are used to determine systematic uncertainties in the $B_s^0$ fit. These fit variations are described in Section \ref{sec:Systematics}.

\subsection{Muon trigger proper time-dependent efficiency\label{sec:Teff}}
It was observed that the muon trigger biases the transverse impact parameter of muons, resulting in a minor inefficiency at large values of the proper decay time.
 This inefficiency is measured using MC simulated events,  by comparing  the \Bs\ proper decay time distribution  of an unbiased sample with the  distribution obtained including the trigger.   To account for this inefficiency in the fit, the events are re-weighted by a factor $w$:
\begin{equation}
  w =     p_0 \cdot [ 1 - p_1 \cdot (\rm{Erf} ((t-p_3)/p_2) + 1)],
   \label{eq:LifeTimeWeight}
\end{equation}
where  $p_0, p_1, p_2$ and $p_3$ are parameters determined in the fit to MC events. No significant bias or inefficiency due to off-line track reconstruction, vertex reconstruction, or track quality selection criteria is observed.

%% file: Section6.tex
\section{Results\label{sec:results}}
The full simultaneous unbinned maximum-likelihood fit contains  nine physical parameters: \DGs, \phis, \Gs, $|A_{0}(0)|^2$, $|A_{\parallel}(0)|^2$, $\delta_{||}$, $\delta_{\perp}$, $|A_{S}(0)|^2$ and $\delta_{S}$.  The other parameters in the likelihood function are the \Bs\ signal fraction $f_s$, parameters describing the $J/\psi \phi$ mass distribution, parameters describing the \Bs\ meson decay time plus angular distributions of background events, parameters used to describe the estimated decay time uncertainty distributions for signal and background events, and scale factors between the estimated decay time  uncertainties and their true uncertainties.  In addition there are also 353 nuisance parameters describing the background and acceptance functions that are fixed at the time of the fit. The fit model is tested using pseudo-experiments as described in Section~\ref{sec:Systematics}. These tests show no significant bias, as well as no systematic underestimation of the statistical errors reported from the fit to data.

\begin{figure}[htbp]
\begin{center}
\ifthenelse {\boolean{Nothing}}
{ \includegraphics[width=0.45\textwidth]{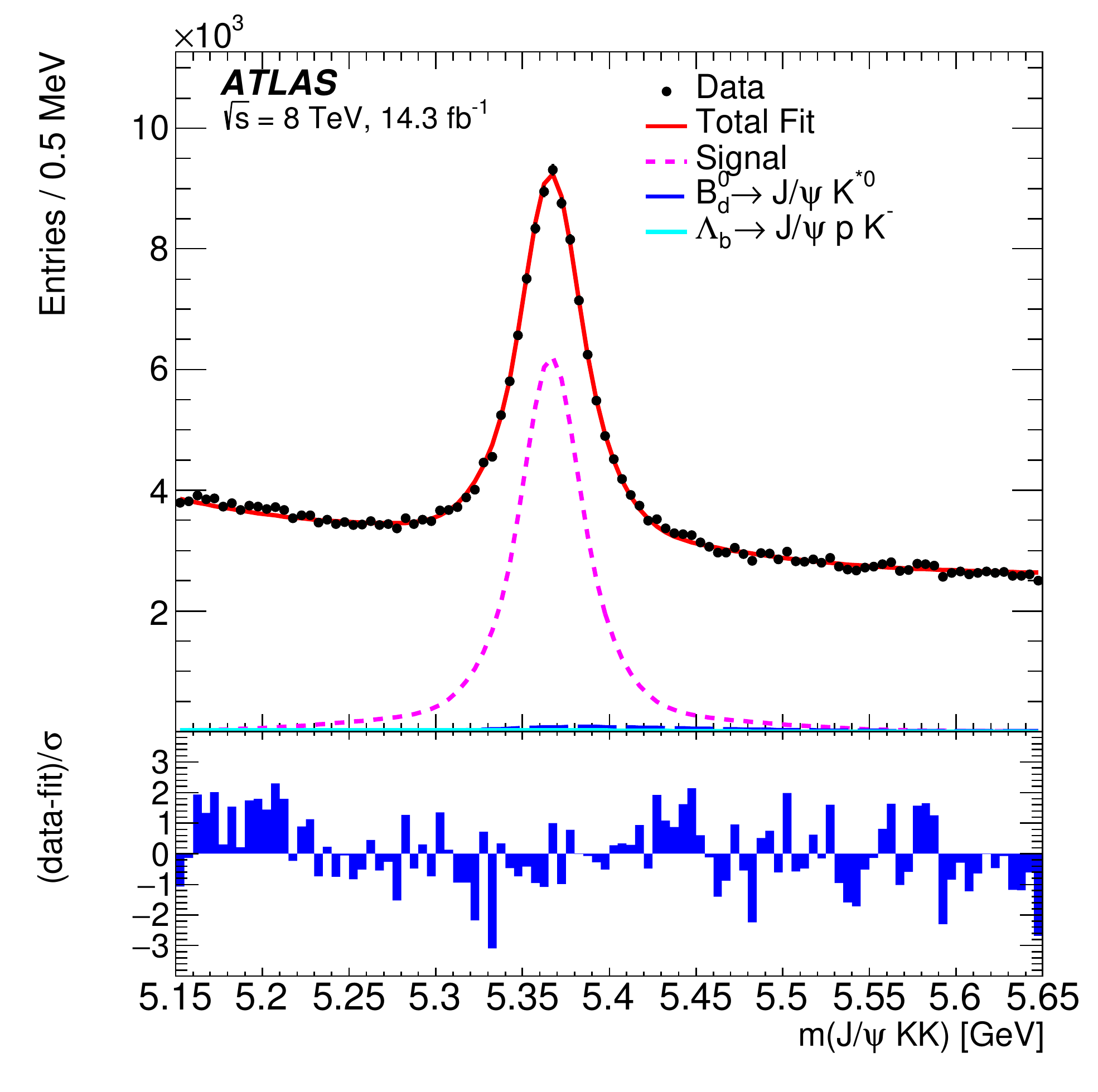}
\includegraphics[width=0.45\textwidth]{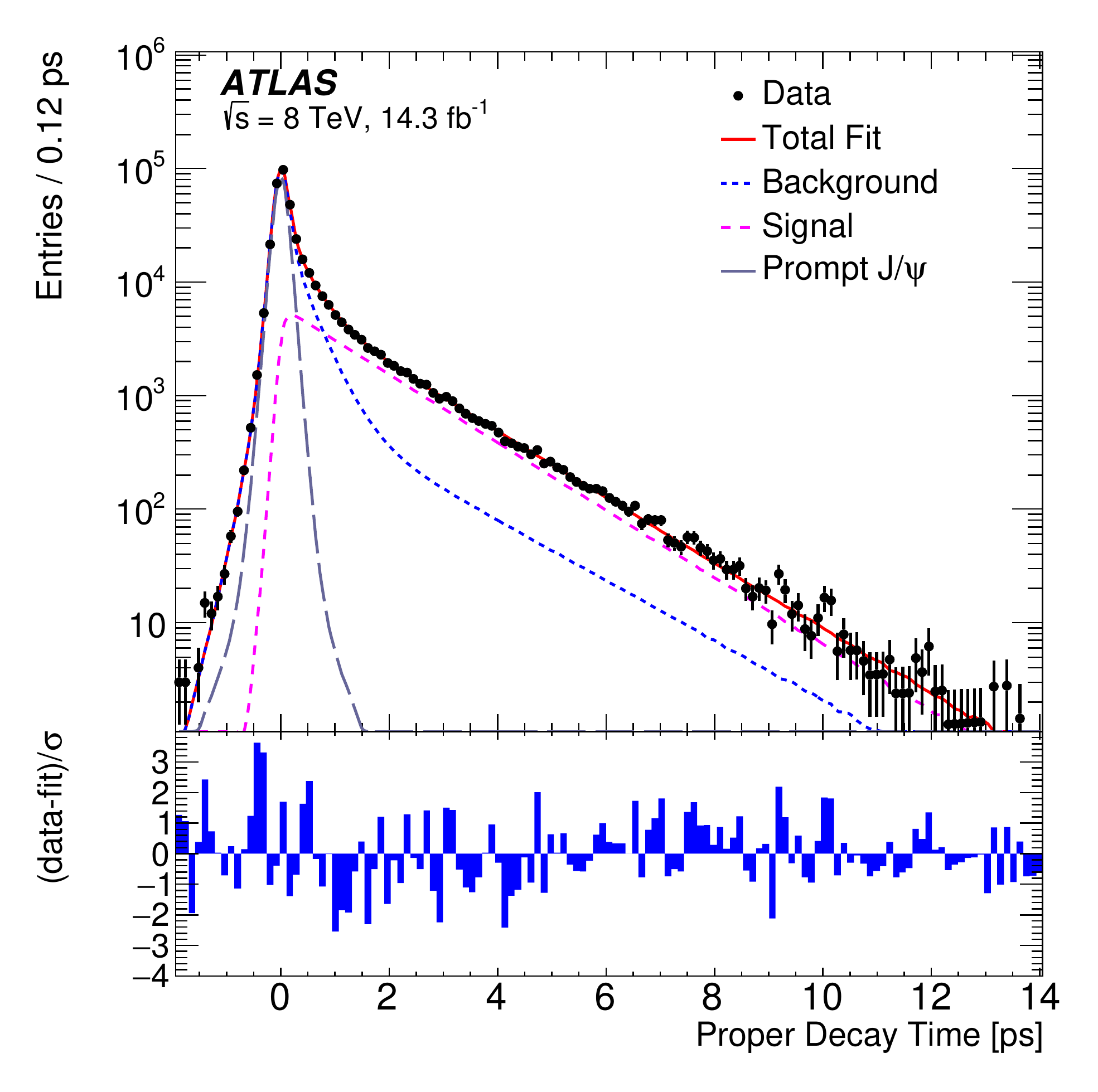} }
{
\ifthenelse {\boolean{Internal}}
{ \includegraphics[width=0.45\textwidth]{mass_Internal.pdf}
\includegraphics[width=0.45\textwidth]{time_Internal.pdf} }
{\includegraphics[width=0.45\textwidth]{mass_Preliminary.pdf}
\includegraphics[width=0.45\textwidth]{time_Preliminary.pdf} }
}
\caption{(Left)  Mass fit projection for the \Bst \ sample.  The red line shows the total fit, the dashed purple line shows the signal component, the long-dashed dark blue line shows the $\Bd \ra J/\psi K^{0*}$ component, while the solid light blue line shows the contribution from $\Lambda_b \ra J/\psi pK^{-}$ events. (Right) Proper decay time fit projection for the \Bst \ sample. The red line shows the total fit while the purple dashed line shows the total signal.  
The total background is shown as a blue dashed line with a long-dashed grey line showing the prompt $J/\psi$ background. Below each figure is a ratio plot that shows the difference between each data point and the total fit line divided by the statistical uncertainty ($\sigma$) of that point.}

\label{fig:simulFit}
\end{center}
\end{figure}

\begin{figure}[htbp]
\begin{center}
\ifthenelse {\boolean{Nothing}}
{\includegraphics[width=0.45\textwidth]{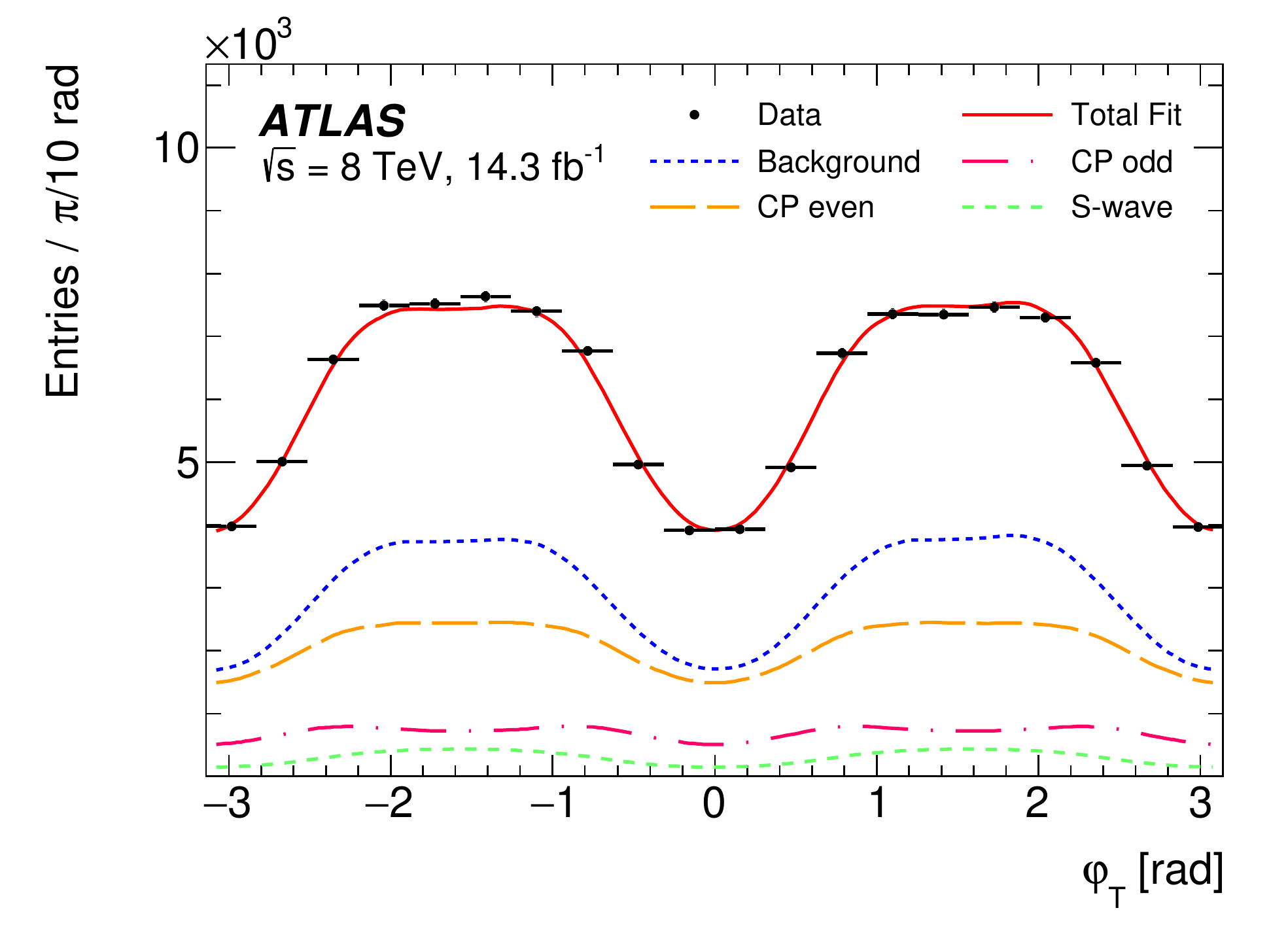}
\includegraphics[width=0.45\textwidth]{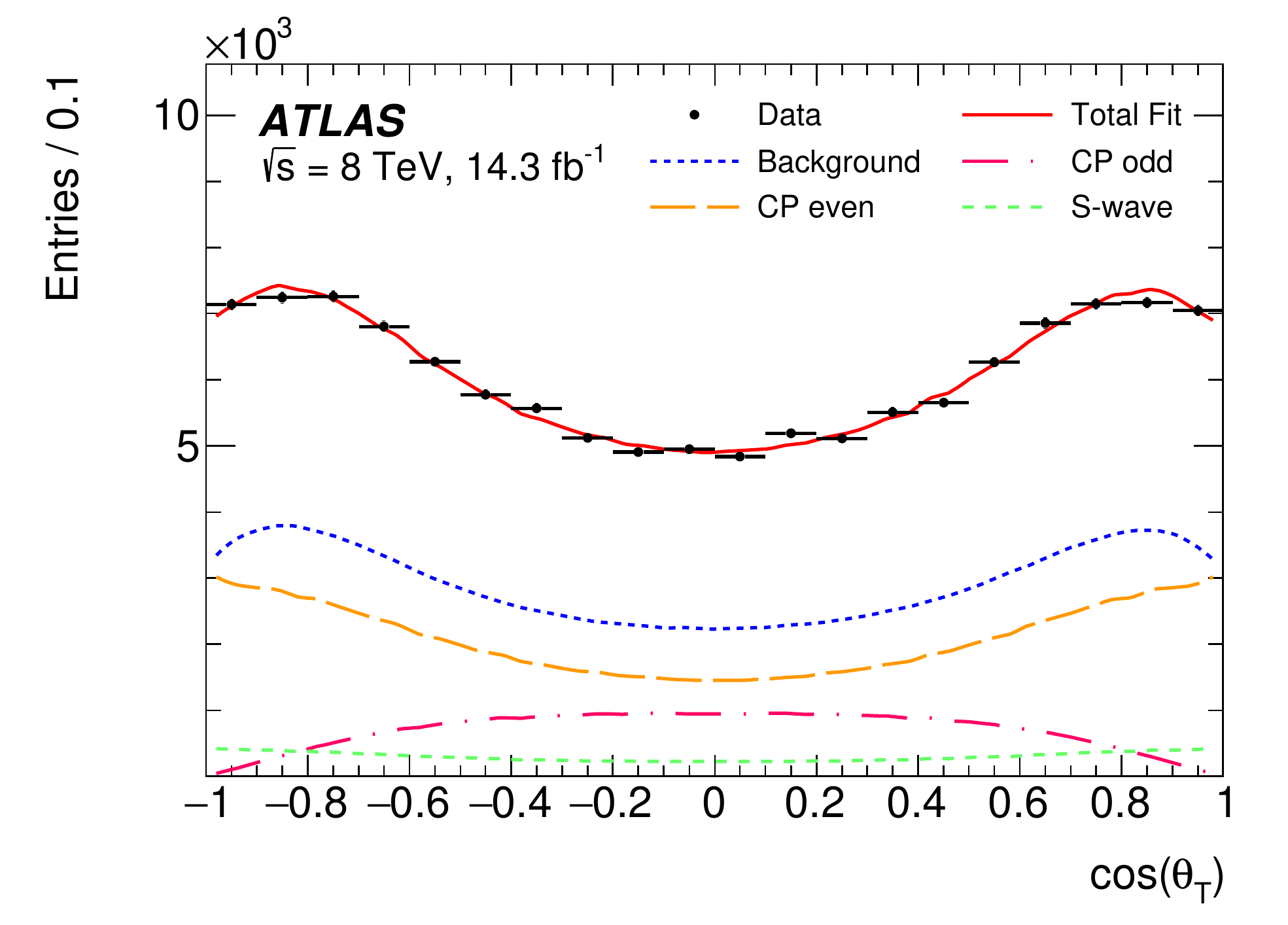}
\includegraphics[width=0.45\textwidth]{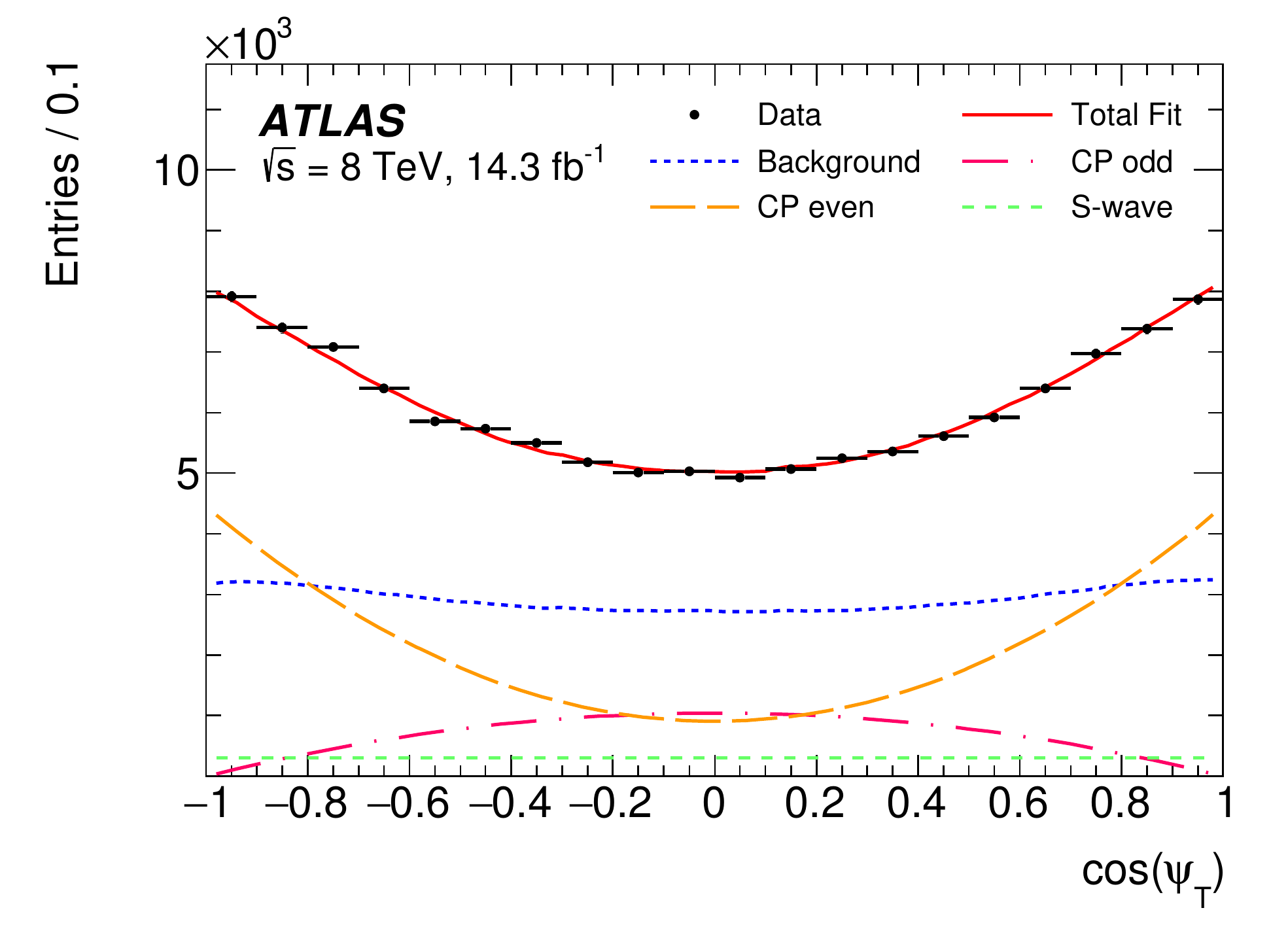}}
{
\ifthenelse {\boolean{Internal}}
{\includegraphics[width=0.45\textwidth]{phi_Internal.pdf}
\includegraphics[width=0.45\textwidth]{costheta_Internal.pdf}
\includegraphics[width=0.45\textwidth]{cospsi_Internal.pdf}}
{\includegraphics[width=0.45\textwidth]{phi_Preliminary.pdf}
\includegraphics[width=0.45\textwidth]{costheta_Preliminary.pdf}
\includegraphics[width=0.45\textwidth]{cospsi_Preliminary.pdf}}
}

\caption{Fit projections for the transversity angles of events with $5.317$ \GeV\ $< m(J/\psi KK) < 5.417$ \GeV\ for $\phi_T$ (top left), $\cos(\theta_T)$ (top right), and $\cos(\psi_T)$ (bottom).  In all three plots the red solid line shows the total fit, the CP-odd and CP-even signal components are shown by the red dot--dashed and orange dashed lines respectively, the S-wave component is given by the green dashed line and the blue dotted line shows the background contribution. The contributions of the interference terms are negligible in these projections and are not shown.}
\label{fig:AngularProjections}
\end{center}
\end{figure}

Multiplying the total number of events supplied to the fit with the extracted signal fraction and its statistical uncertainty provides an estimate for the total number of \Bs\ meson candidates of $74900 \pm 400$. The results and correlations of the physics parameters obtained from the fit are given in Tables~\ref{tab:FitResults} and~\ref{tab:FitResultsCor}.  Fit projections of the mass, proper decay time and angles are given in Figures~\ref{fig:simulFit} and \ref{fig:AngularProjections}, respectively.  

\begin{table}[t]
\begin{center}
\begin{tabular}{c|c|c|c}
\hline
Parameter & Value & Statistical & Systematic \\
 & & uncertainty & uncertainty \\
\hline\hline
 \phis [rad] & \pSfit & \pSstat & \pSsyst \\
 
 \DGs [ps$^{-1}$] & \DGfit & \DGstat & \DGsyst \\
 \Gs [ps$^{-1}$] & \GSfit & \GSstat & \GSsyst \\
  $|A_{\parallel}(0)|^2$ & \Apasqfit & \Apasqstat & \Apasqsyst \\
  $|A_{0}(0)|^2$ & \Azesqfit  & \Azesqstat & \Azesqsyst \\
$|A_{S}(0)|^2$ & \ASsqfit & \ASsqstat & \ASsqsyst \\
$\delta_\perp$ [rad]& \Deltperpfit  & \Deltperpstat & \Deltperpsyst \\
$\delta_{\parallel}$ [rad]  & \Deltparafit & \Deltparastat &\Deltparasyst \\
$\delta_{\perp} -\delta_{S}$ [rad] & \DeltPSfit  & \DeltSstat & \DeltPSsyst  \\

\hline
\end{tabular}
\end{center}
\caption{Fitted values for the physical parameters of interest with their statistical and systematic uncertainties. \label{tab:FitResults} }
\end{table}

\begin{table}[htp]
\begin{center}
\begin{tabular}{|c|c|c|c|c|c|c|c|c|c|c|}
\hline
			& $\Delta\Gamma$	& $\Gamma_{s}$ 	& $|A_{||}(0)|^2$ 	& $|A_0(0)|^2$ 	& $|A_S(0)|^2$ 	& $\delta_{\parallel}$  & $\delta_{\perp}$  & $\delta_{\perp} -\delta_{S} $ \\	
\hline
$\phi_{s}$ 		& \pDG		& \pG 	& \phantom{-}\pApa 	 	& \phantom{-}\pAz 	& \phantom{-}\pAs 	& \phantom{-}\pDpara     & \phantom{-}\pDperp    &  \pDpS  \\
$\Delta\Gamma$ 	& 1 			& \DGG	& \phantom{-}\DGApa 	& \phantom{-}\DGAz 	& \phantom{-}\DGAs 	& \phantom{-}\DGDpara  & \phantom{-}\DGDperp & \DGDpS \\
$\Gamma_{s}$ 		&  			& 1 		& \GApa     			& \GAz 			& \phantom{-}\GAs 	& \GDpara    			& \GDperp  			 & \phantom{-}\GDpS \\
 $|A_{||}(0)|^2$ 	        &		  	&  		& 1	 				&  \AzApa 		& \phantom{-}\ApaAs 	& \phantom{-}\ApaDpara  & \phantom{-}\ApaDperp  & \ApaDpS \\
 $|A_0(0)|^2$ 		&  			&  		&  					& 1	 			& \phantom{-}\AzAs 	& \AzDpara 			 & \phantom{-}\AzDperp & \phantom{-}\AzDpS \\
  $|A_S(0)|^2$ 		&  			&  		&  					&  				& 1	 			& \AsDpara  			& \AsDperp 			& \phantom{-}\AsDpS\\
   $\delta_{\parallel}$ &  			&  		&  					&  				& 	 			& 1		 			 & \phantom{-}\DparaDperp & \DparaDpS \\
 $\delta_{\perp}$       &  			&  		&  					&  				& 	 			& 					& 1	   &  \phantom{-}\DperpDpS \\
 \hline
\end{tabular}
\end{center}
\caption{Fit correlations between the physical parameters of interest.   \label{tab:FitResultsCor}}
\end{table}%

%% file: Section7.tex
\section{Systematic uncertainties} \label{sec:Systematics}
Systematic uncertainties are assigned by considering effects that are not accounted for in the likelihood fit. These are described below.  
\begin{itemize}

\item {\bf Flavour tagging:} There are two contributions to the uncertainties in the fit parameters due to the flavour tagging procedure, the statistical and systematic components. The statistical uncertainty due to the size of the sample of $B^\pm \ra J/\psi K^\pm$ decays is included in the overall statistical error.
The systematic uncertainty arising from the precision of the tagging calibration is estimated by changing the model used to parameterize the probability distribution, $P(B|Q)$, as a function of tag charge from the third-order polynomial function used by default to one of several alternative functions. The alternatives used are: a linear function; a fifth-order polynomial; or two third-order polynomials describing the positive and negative regions that share the constant and linear terms but have independent quadratic and cubic terms. For the combined-muon tagging, an additional model consisting of two third-order polynomials sharing the constant term but with independent linear, quadratic and cubic terms is also used.
The \Bs\ fit is  repeated using the alternative models and the largest difference is assigned as the systematic uncertainty.

\item {\bf Angular acceptance method:} The angular acceptance (from the detector and kinematic effects mentioned in Section \ref{sec:sigPDF}) is calculated from a binned fit to MC simulated data.  In order to estimate the size of the systematic uncertainty introduced from the choice of binning, different acceptance functions are calculated using different bin widths and central values. These effects are found to be negligible. 

\item {\bf Inner detector alignment:} Residual misalignments of the ID affect the impact parameter, $d_0$, distribution with respect to the primary vertex. The effect of a radial expansion on the measured $d_0$ is determined from data collected at   \CoMEnergy, with a trigger requirement of at least one muon with a transverse momentum greater than or equal to $4$ \GeV. The radial expansion uncertainties determined in this way are 0.14\% for  $|\eta|<1.5$ and 0.55\%  for $1.5<|\eta|<2.5$. These values are used to estimate the effect on the fitted \Bs\  parameter values.Ê Small deviations are seen in some parameters, and these are included as systematic uncertainties.

\item {\bf Trigger efficiency:}  To correct for the trigger lifetime bias the events are re-weighted according to Equation~(\ref{eq:LifeTimeWeight}). The uncertainty of the parameters  $ p_0, p_1, p_2$ and $p_3$  are used to estimate the systematic uncertainty due to the time efficiency correction. These uncertainties originate from the following sources:  the limited size of the MC simulated dataset, the choice of bin-size for the proper decay time distributions and variations between different triggers.  The  systematic effects are found to be negligible.

\item  {\bf Background angles model, choice of \boldmath$p_\mathrm{T}$ bins:} The shape of the background angular distribution, $P_\textrm{b}(\theta_T, \varphi_T, \psi_T)$, is described by the Legendre polynomial functions given in  Equation (\ref{eq:likelihoodBkgAng}).  The shapes arise primarily from detector and kinematic acceptance effects and are sensitive to the \psubt \ of the \Bs \ meson candidate. For this reason, the parameterization using the Legendre polynomial functions is performed in four \psubt \ intervals: 0--13 \GeV, 13--18 \GeV, 18--25 \GeV \ and $>$25 \GeV. The systematic uncertainties due to the choice of  \psubt intervals are estimated by repeating the fit, varying  these intervals. The biggest deviations observed in the fit results were taken to represent the systematic uncertainties. 

\item  {\bf Background angles model, choice of mass sidebands:} 
 The parameters of the Legendre polynomial functions given in  Equation (\ref{eq:likelihoodBkgAng}) are adjusted to give the best fit to the angular distributions for events in the \Bs\ mass sidebands.  To test the sensitivity of the fit results to the choice of sideband regions, the fit is repeated with alternative  choices for the excluded signal mass regions: $|m(B_s^0) - 5.366|> 0.085$ \GeV\ and $|m(B_s^0) - 5.366|> 0.160$ \GeV\ (instead of $|m(B_s^0) - 5.366|> 0.110$ \GeV). The differences in the fit results are assigned as systematic uncertainties.

\item  {\bf \boldmath{$B_d$}  contribution:} The contamination from $B_d \ra J/\psi K^{0*}$ events mis-reconstructed as \Bst\ is accounted for in the final fit. Studies are performed to evaluate the effect of the uncertainties in the $B_d \ra J/\psi K^{0*}$  fraction,  and the shapes of the mass and transversity angles distribution.  In the MC events the angular distribution of the $B_d \ra J/\psi K^{0*}$ decay is modelled using parameters taken from  Ref.~\cite{LHCbK0Star}. The uncertainties of these parameters are taken into account in the estimation of systematic uncertainty.  After  applying the \Bs \ signal selection cuts,   the angular distributions are fitted  using Legendre polynomial functions. The  uncertainties of this fit are included in the  systematic tests.  The impact of all these uncertainties is found to have a negligible effect on the \Bs\  fit results.  The contribution of $B_d \ra J/\psi K\pi$ events as well as their interference with $B_d \ra J/\psi K^{0*}$ events is not included in the fit and is instead assigned as a systematic uncertainty. To evaluate this uncertainty, the MC background events are modelled using both the P-wave $B_d \ra J/\psi K^{0*}$ and S-wave $B_d \ra J/\psi K\pi$ decays and their interference, using the input parameters taken from Ref.~\cite{LHCbK0Star}. The  \Bs\  fit  using this input was compared to the default fit, and differences are included in Table~\ref{tab:syst_totals}. 

\item  {\bf \boldmath{$\Lambda_b$}  contribution:} The contamination from $\Lambda_b \ra J/\psi p K^{-}$ events mis-reconstructed as \Bst\ is accounted for in the final fit. Studies are performed to evaluate the effect of the uncertainties in the $\Lambda_b \ra J/\psi p K^{-}$ fraction $f_{\Lambda_b}$, and the shapes of the mass, transversity angles, and lifetime distributions. Additional studies are performed to determine the effect of the uncertainties in the $\Lambda_b \ra J/\psi \Lambda^{*}$ branching ratios used to reweight the generated MC. These are uncertainties are included in Table~\ref{tab:syst_totals}.

\item {\bf  Fit model variations} To estimate the systematic uncertainties due to the fit model, variations of the model are tested in pseudo-experiments. A set of $\approx$2500 pseudo-experiments is generated for each variation considered, and fitted with the default model.   The systematic error quoted for each effect is the difference between the mean shift of the fitted value of each parameter from its input value for the pseudo-experiments altered for each source of systematic uncertainty. In the  first variation  tested, the signal mass is generated using  the fitted \Bs \ mass  convolved  with a Gaussian function using  the measured  per-candidate mass errors. In another test, the background mass is generated from an exponential function with the addition of a first-degree polynomial function instead of an exponential function plus a constant term.   
The time resolution model was varied by using two different scale factors to generate the lifetime uncertainty, instead of the single scale factor used in the default model. The non-negligible uncertainties derived from these tests are included in the systematic uncertainties shown in Table~\ref{tab:syst_totals}. 
To determine the possible systematics effects of mis-modelling  of the background events  by the fitted background model, as seen in the low mass side-band region (5.150--5.210~GeV) of Figure \ref{fig:simulFit}, left,  alternative mass selection cuts are used with the default fit model. The effect of these changes on the fit results are found to be negligible.

\item {\bf Default fit model:} Due to its complexity, the fit model  is less sensitive to some   nuisance parameters. This limited sensitivity  could potentially lead to a bias in the measured physics parameters, even when the model perfectly describes the fitted data. To estimate the systematic uncertainty due to the choice of default fit model,  a set of pseudo-experiments were conducted using the default model in both the generation and fit.  The systematic uncertainties are determined from the mean of the pull distributions of the pseudo-experiments scaled by the statistical error of that parameter on the fit to data.  These tests show no significant bias in the fit model, and no systematic underestimation of the statistical errors reported from the fit to data.

\end{itemize}
The systematic uncertainties are listed in Table~\ref{tab:syst_totals}. For each parameter, the total systematic error is obtained by adding  all of the contributions in quadrature.

\begin{table}[ht]
\footnotesize
\begin{center}
\begin{tabular}{l| ccc ccc ccc} 
 \hline \hline 
 & \phis  & \DGs & \Gs   &  $|A_{\parallel}(0)|^2$ & $|A_{0}(0)|^2$ & $ |A_{S}(0)|^2 $ &  $\delta_{\perp}$& $\delta_{\parallel}$  & $\delta_{\perp} -\delta_{S} $\\ 
&[rad]&[ps$^{-1}$]&[ps$^{-1}$]& &&& [rad]&[rad]&  [rad] \\
\hline
&&& &&& &&&  \\
\ifthenelse {\boolean{BdFraction3.3}}
{Tagging                      & 0.025    & 0.003    & \sma    & \sma    & \sma    & 0.001   & 0.236    & 0.014    & 0.004 \\}
 {Tagging                   & 0.026    & 0.003    & \sma    & \sma    & \sma    & 0.001   & 0.238    & 0.014    & 0.004 \\}

Acceptance                 & \sma    & \sma    & \sma    &  0.003    & \sma    & 0.001    & 0.004    & 0.008    & \sma  \\
Inner detector alignment & 0.005    & \sma    & 0.002    &  \sma     & \sma    & \sma     & 0.134    & 0.007  & \sma  \\

Background angles model: &&& &&& &&&  \\
\hspace{0.5cm}Choice of \psubt bins           & 0.020      & 0.006    & 0.003    & 0.003    & \sma    & 0.008    & 0.004    & 0.006    & 0.008 \\
\hspace{0.5cm}Choice of mass interval    & 0.008    & 0.001    & 0.001    & \sma    & \sma    & 0.002    & 0.021    & 0.005    & 0.003 \\
\ifthenelse {\boolean{BdFraction3.3}}
{\Bd \  background  model  &  0.023     &  0.001  &  \sma   &  0.002  &  0.002 &  0.017   &  0.090  & 0.011 & 0.009  \\ }
{\Bd \  background  model  &  0.008     &  \sma  &  \sma   &  0.001  &  \sma &  0.008   &  0.007  & \sma & 0.005  \\ }

$\Lambda_b$ background  model  &  0.011     &  0.002  &  0.001   &  0.001  &  0.007  &  0.009   &  0.045  & 0.006 & 0.007  \\

Fit model: &&& &&& &&&  \\
\hspace{0.5cm}Mass signal model     & 0.004  & \sma & \sma & 0.002 & \sma & 0.001 & 0.015 & 0.017 & \sma \\
\hspace{0.5cm}Mass background model & \sma & 0.002   & \sma & 0.002 & \sma & 0.002   & 0.027 & 0.038 & \sma \\
\hspace{0.5cm}Time resolution model & 0.003  & \sma  & 0.001  & 0.002 & \sma & 0.002  & 0.057 & 0.011 & 0.001  \\
\hspace{0.5cm}Default fit  model       & 0.001  & 0.002  & \sma  & 0.002 & \sma & 0.002  & 0.025 & 0.015 & 0.002  \\
 \hline 
\hline      
          &&& &&& &&&  \\       
\bf Total  & \pSsyst & \DGsyst & \GSsyst  & \Apasqsyst & \Azesqsyst  & \ASsqsyst & \Deltperpsyst & \Deltparasyst & \DeltPSsyst \\
\hline \hline
\end{tabular}
\end{center}
\caption{Summary of systematic uncertainties assigned to the physical parameters of interest.}
\label{tab:syst_totals}
\end{table}

%% file: Sections8-9.tex
\section{Discussion}
\begin{figure}[!htp]
\begin{center}
\ifthenelse {\boolean{Nothing}}
{\includegraphics[width=0.32\textwidth]{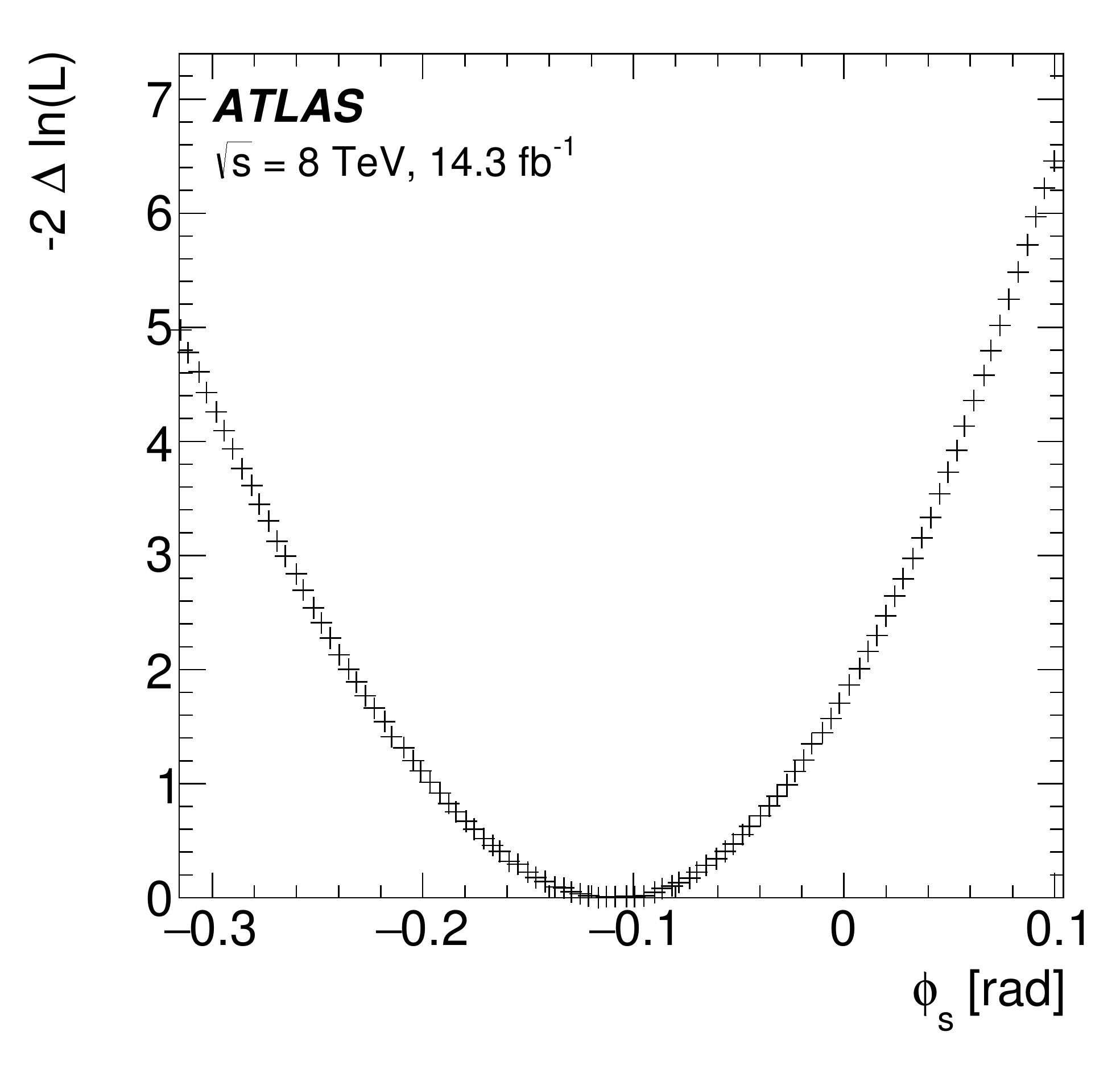}
\includegraphics[width=0.32\textwidth]{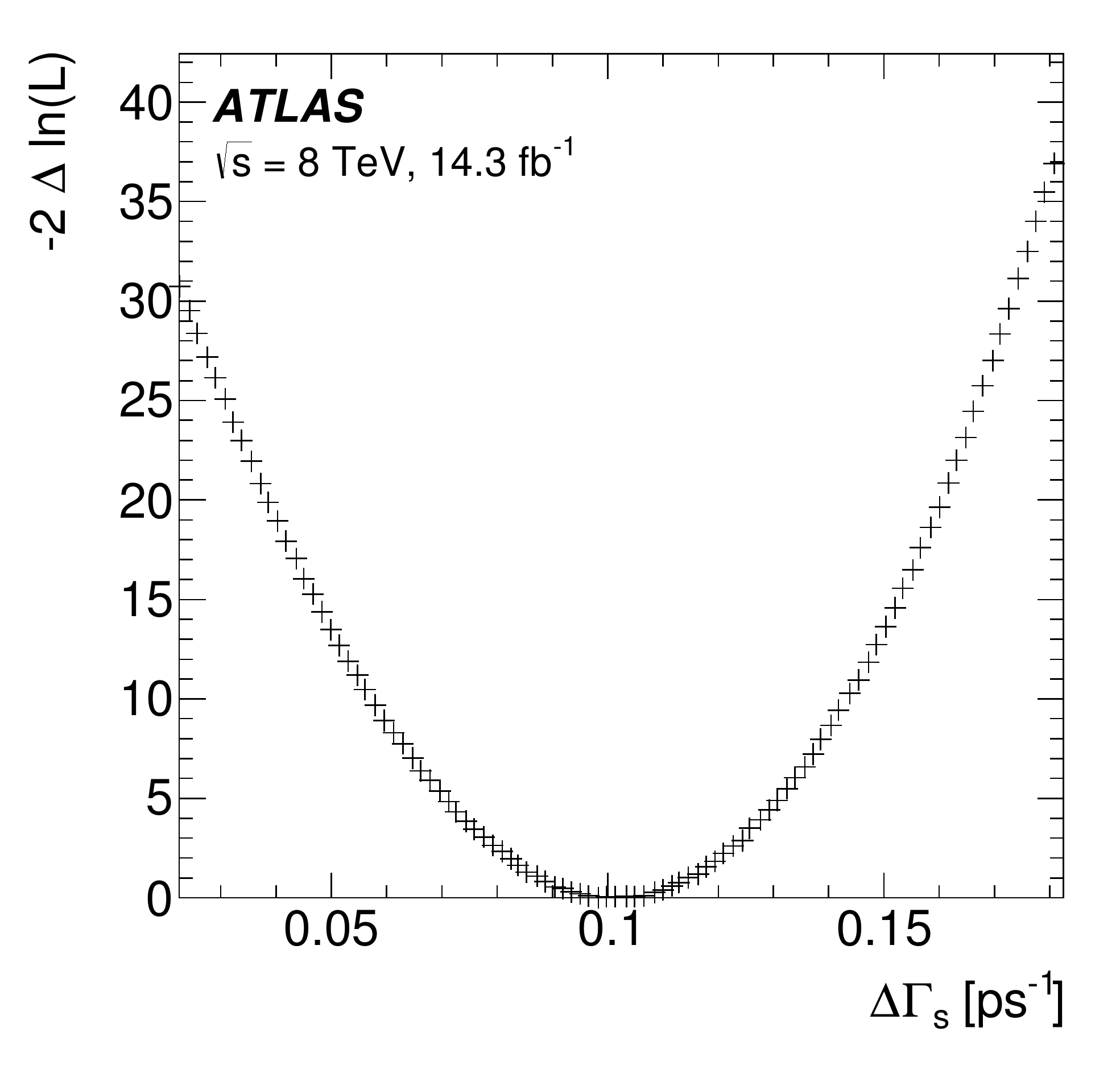}
\includegraphics[width=0.32\textwidth]{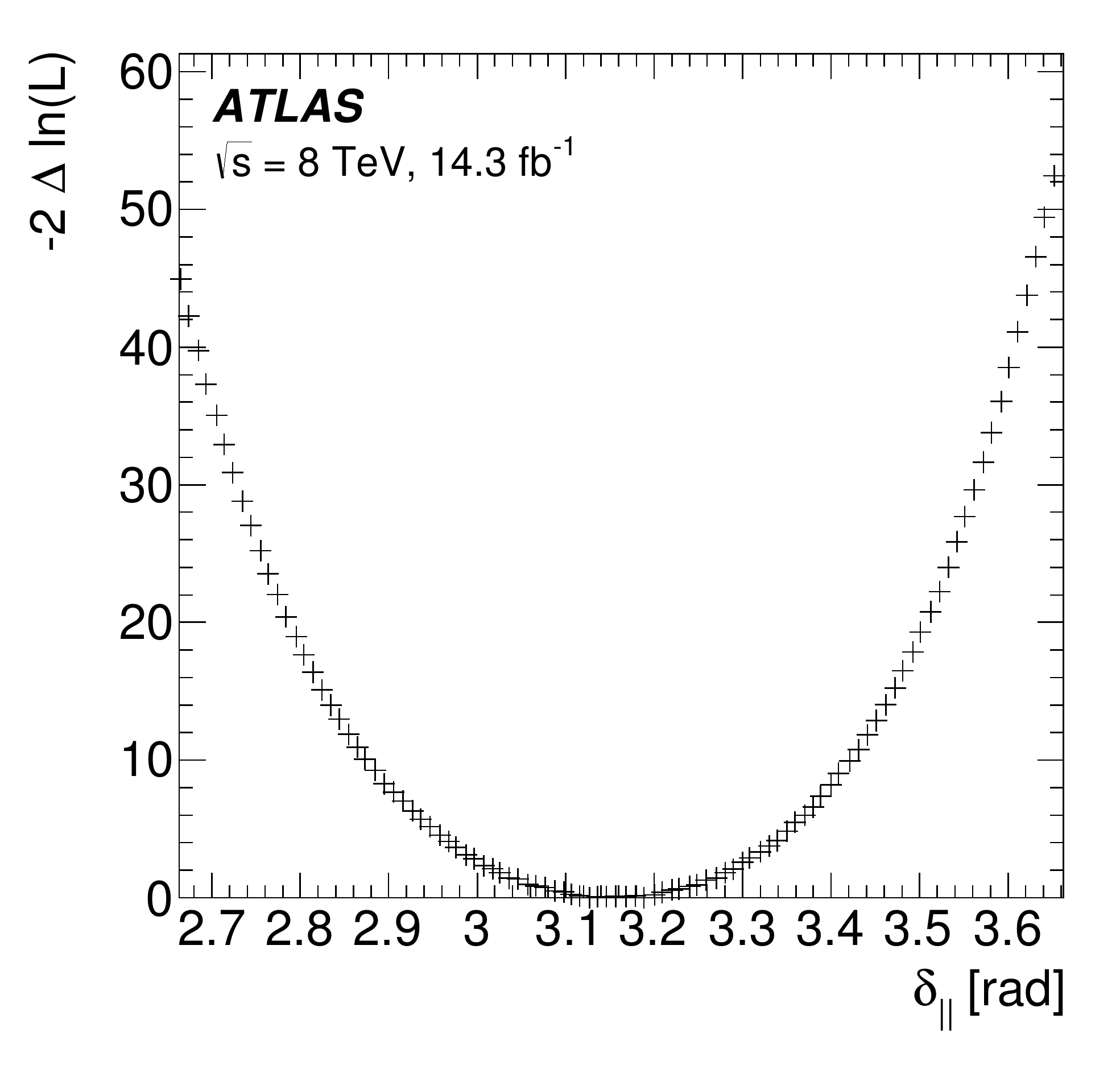}
\includegraphics[width=0.32\textwidth]{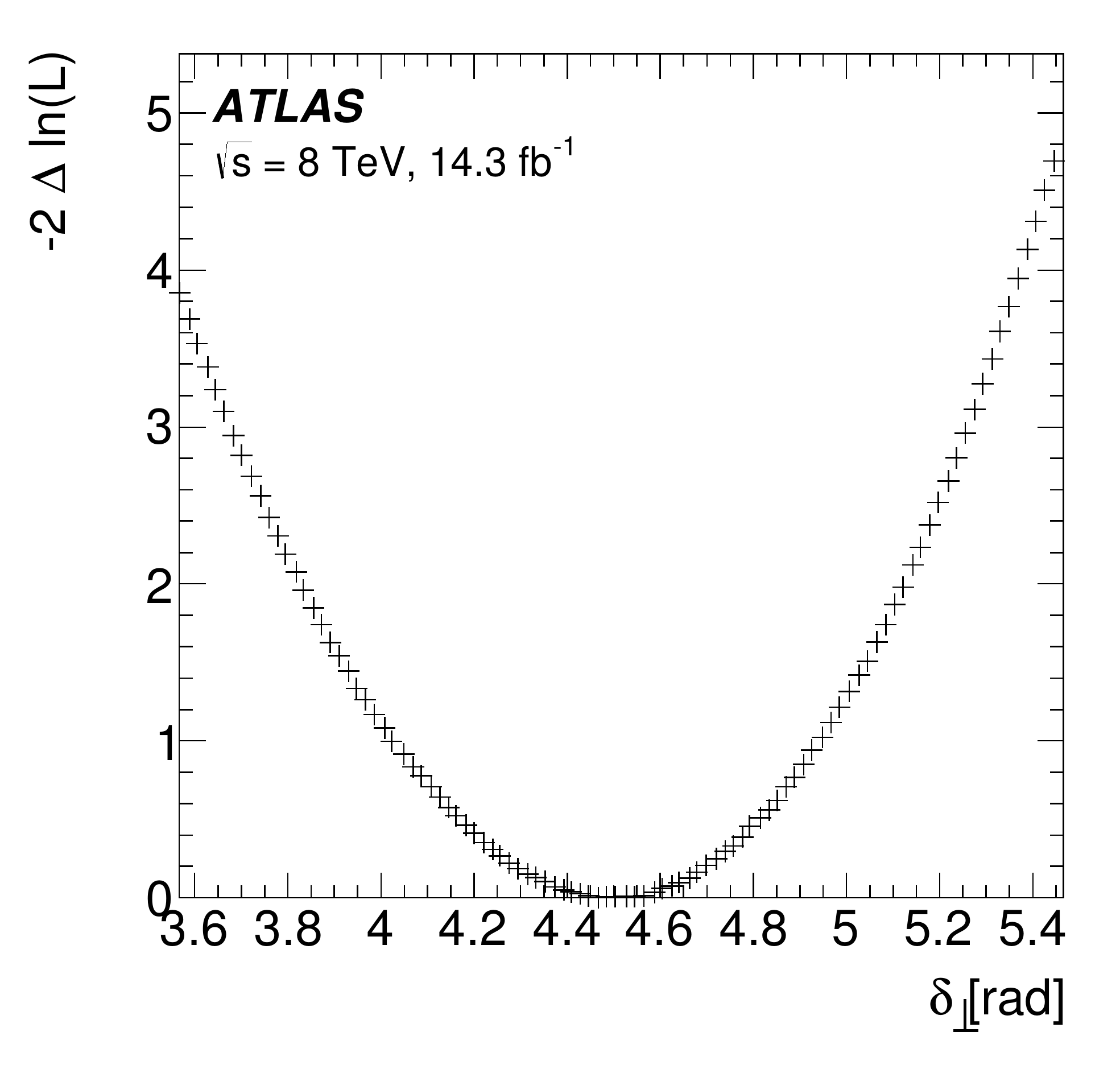}
\includegraphics[width=0.32\textwidth]{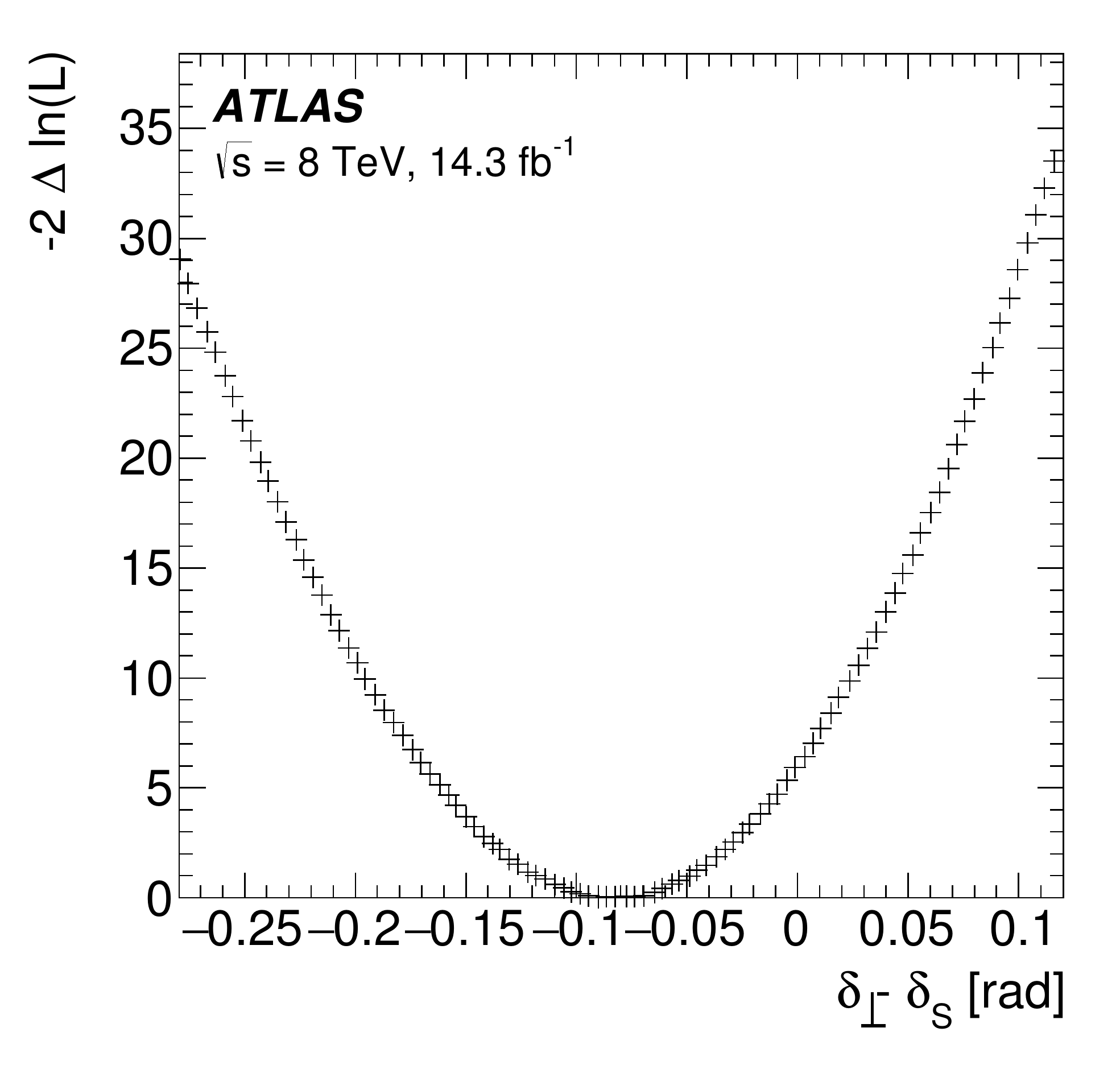}}
{
\ifthenelse {\boolean{Internal}}
{\includegraphics[width=0.32\textwidth]{lnL_phiS_Internal.pdf}
\includegraphics[width=0.32\textwidth]{lnL_DeltaGamma_Internal.pdf}
\includegraphics[width=0.32\textwidth]{lnL_deltaPara_Internal.pdf}
\includegraphics[width=0.32\textwidth]{lnL_deltaPerp_Internal.pdf}
\includegraphics[width=0.32\textwidth]{lnL_deltaS_perp_Internal.pdf}}
{\includegraphics[width=0.32\textwidth]{lnL_phiS_Preliminary.pdf}
\includegraphics[width=0.32\textwidth]{lnL_DeltaGamma_Preliminary.pdf}
\includegraphics[width=0.32\textwidth]{lnL_deltaPara_Preliminary.pdf}
\includegraphics[width=0.32\textwidth]{lnL_deltaPerp_Preliminary.pdf}
\includegraphics[width=0.32\textwidth]{lnL_deltaS_perp_Preliminary.pdf}}
}

\caption{1D likelihood contours (statistical errors only) for  \phis\ (top left), \DGs\  (top centre),  $\delta_{||}$ (top right), $\delta_{\perp}$ (bottom left) and $\delta_\perp - \delta_S$ (bottom right). }
\label{fig:phiSDeltaGLikelihoodScans}
\end{center}
\end{figure}

\begin{figure}[tp]
\begin{center}
\ifthenelse {\boolean{Nothing}}
{\includegraphics[width=0.45\textwidth]{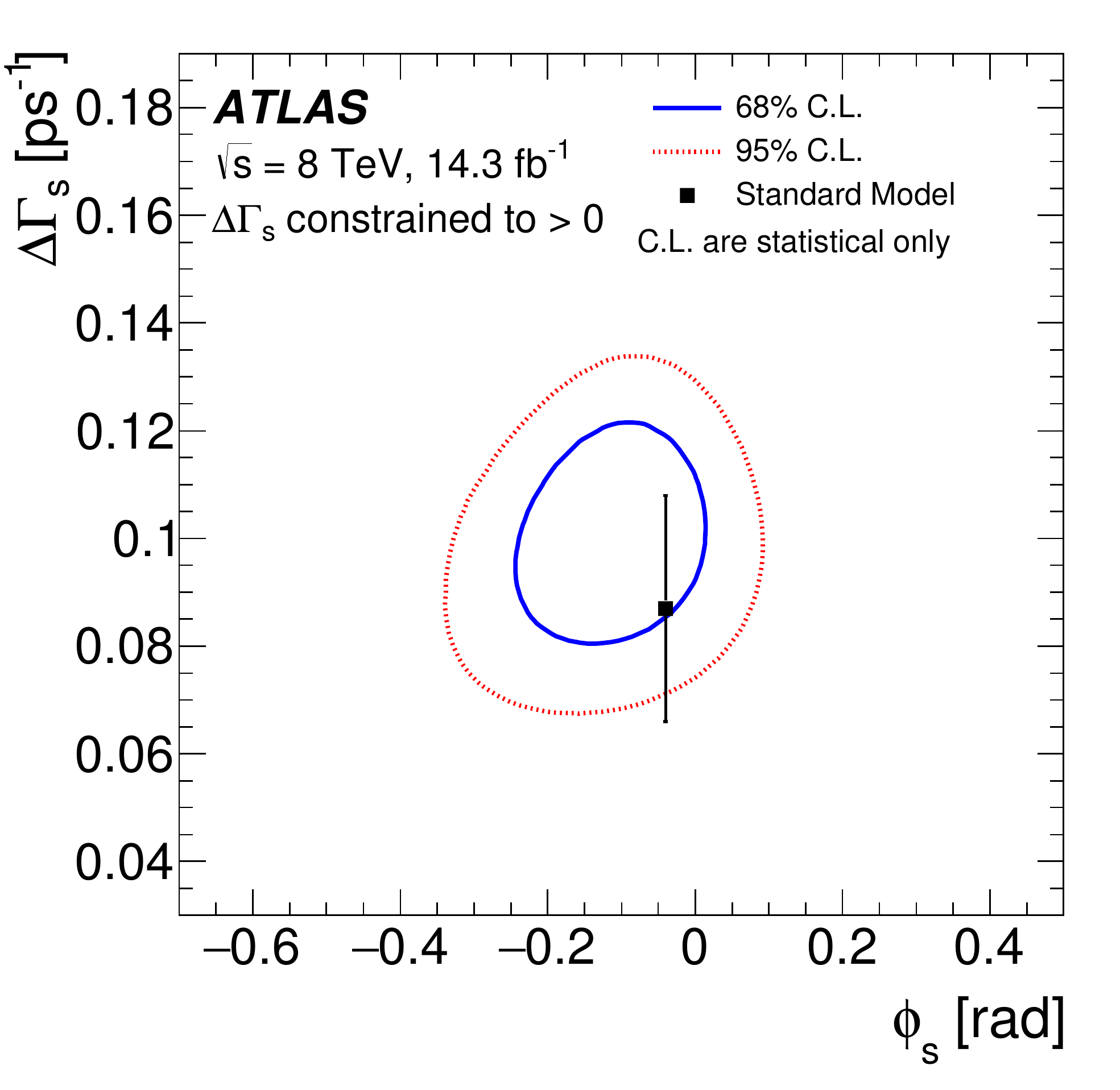}}
{
\ifthenelse {\boolean{Internal}}
{\includegraphics[width=0.45\textwidth]{c_plot_2D_scan_2012_Internal.pdf}}
{\includegraphics[width=0.45\textwidth]{c_plot_2D_scan_2012_Preliminary.pdf}}
}

\caption{Likelihood contours in the \phis--\DGs\ plane for \CoMEnergy\ data.   The blue line shows the 68\% likelihood contour, while the red dotted line shows the 95\% likelihood contour (statistical errors only). The SM prediction is taken from Ref.~\cite{PhysRevD.84.033005}, at this scale the  uncertainty on \phis\ is not visible on the figure.}
\label{fig:Contour2012}
\end{center}
\end{figure}

The PDF describing the \Bst\ decay is invariant under the following simultaneous transformations: 
\begin{equation}
\{\phis, \DGs, \delta_{\perp}, \delta_{\parallel}\}  \rightarrow \{\pi -\phis, -\DGs, \pi - \delta_{\perp}, 2\pi - \delta_{\parallel}\}. \nonumber
\end{equation}
Since \DGs\ was determined to be positive~\cite{Aaij:2012eq},  there is a unique solution.
Figure \ref{fig:phiSDeltaGLikelihoodScans} shows the 1D log-likelihood scans of \phis,\  \DGs \ and of the three measured strong phases $\delta_{||}$, $\delta_{\perp}$ and $\delta_{\perp}-\delta_S$. The variable on vertical axis, $2 \Delta {\rm ln(L)} \equiv  2 {\rm (ln(L^{G})-ln(L^i))}  $, is a difference between the likelihood values of a default fit, $ \rm {(L^G)}$, and of the fit in which the  physical parameter is fixed to a value shown on horizontal axis,  $ \rm {(L^i)}$. $2 \Delta {\rm ln(L)} = 1$ corresponds to the estimated $1 \sigma$ confidence level.   There are a small asymmetries in the likelihood curves, however at the level of one statistical $\sigma$ these are small compared to the corresponding statistical uncertainties of the physical variables, for which the scan is done.  Therefore symmetric statistical uncertainties are quoted. 
Figure \ref{fig:Contour2012} shows the likelihood contours in the \phis--\DGs\ plane. The region predicted by the Standard Model is also shown. 

\begin{figure}[tp]
\begin{center}
\ifthenelse {\boolean{Nothing}}
{\includegraphics[width=0.45\textwidth]{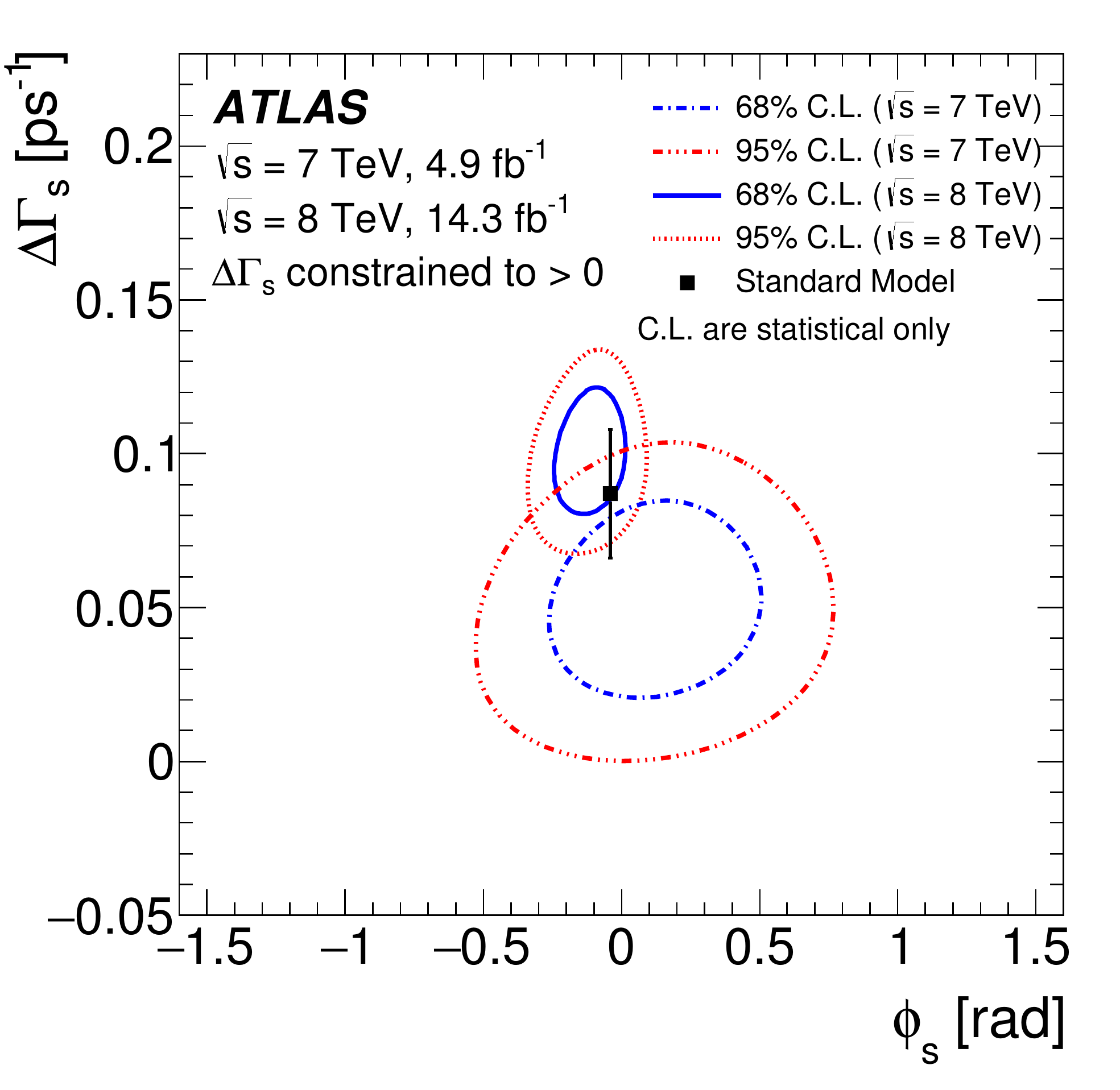}
\includegraphics[width=0.45\textwidth]{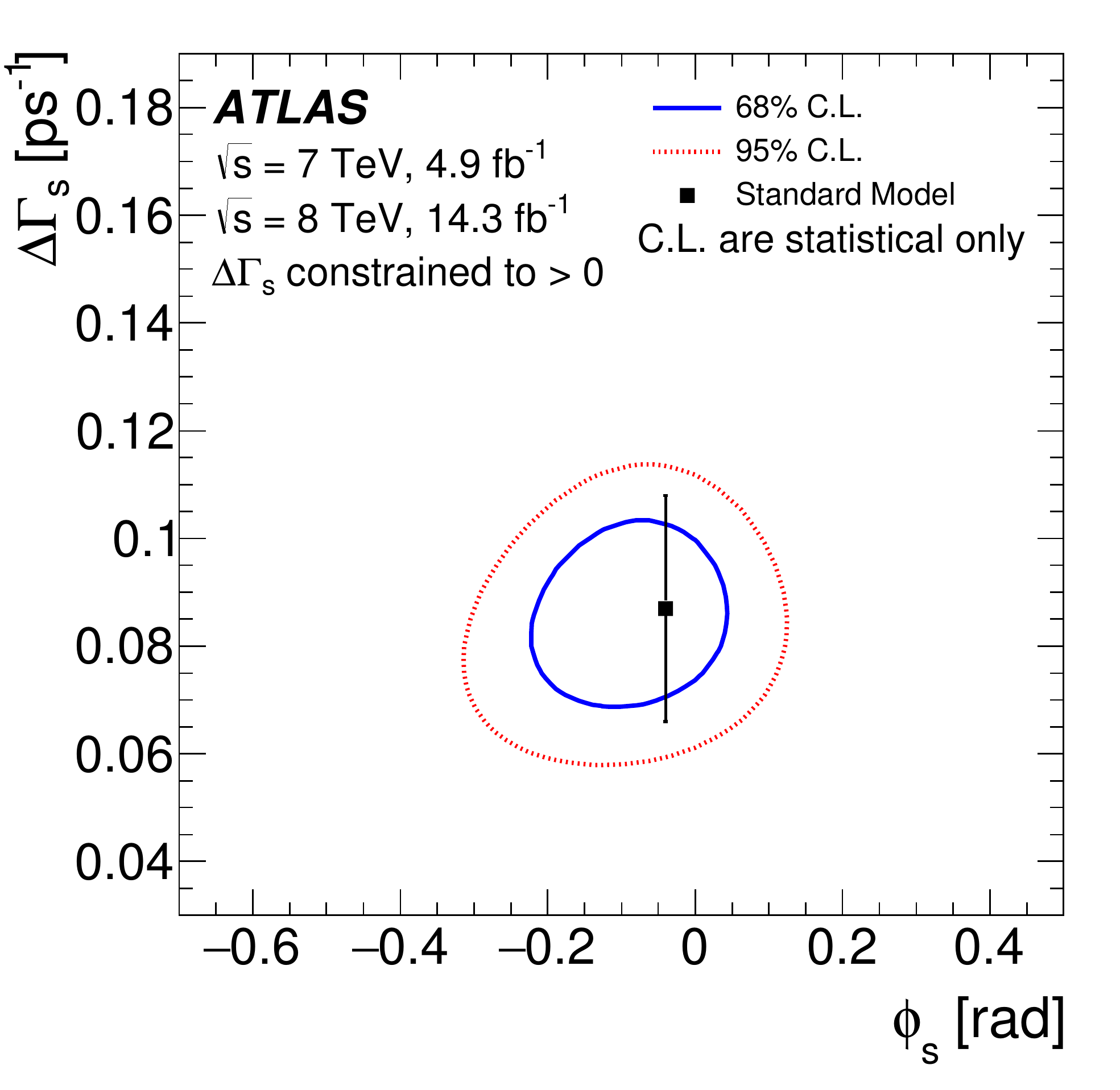}}
{
\ifthenelse {\boolean{Internal}}
{\includegraphics[width=0.45\textwidth]{c_plot_2D_scan_2011and2012_Internal.pdf}
\includegraphics[width=0.45\textwidth]{c_plot_2D_scan_CombinedRun1_Internal.pdf}}
{\includegraphics[width=0.45\textwidth]{c_plot_2D_scan_2011and2012_Preliminary.pdf}
\includegraphics[width=0.45\textwidth]{c_plot_2D_scan_CombinedRun1_Preliminary.pdf}}
}

\caption{Likelihood contours in the \phis--\DGs\ plane for  individual results from \CoMEnergySeven\ and  \CoMEnergy\ data (left) and  a final  statistical combination of the results from \CoMEnergySeven\ and  \CoMEnergy\ data  (right).   The blue line shows the 68\% likelihood contour, while the red dotted line shows the 95\% likelihood contour (statistical errors only). The SM prediction is taken from Ref.~\cite{PhysRevD.84.033005}, at this scale the uncertainty on \phis\ is not visible on the figure.}
\label{fig:Contour}
\end{center}
\end{figure}


\section{Combination of \CoMEnergySeven\ and \CoMEnergy\ results}
The measured values  are consistent with those obtained in a previous analysis  \cite{tagCPV2011}, using ATLAS data collected in 2011 at a centre-of-mass energy of \CoMEnergySeven. This consistency is also clear from a comparison of the likelihood contours in the \phis--\DGs\  projection shown in Figure \ref{fig:Contour}.  A Best Linear Unbiased Estimate (BLUE) combination  \cite{BlueManual}  is used to combine the \CoMEnergySeven\ and \CoMEnergy\  measurements to give an overall result for Run~1. In Ref. \cite{tagCPV2011} the strong phases $\delta_{\parallel}$ and $\delta_{\perp}\text{--}\delta_{S}$ were given as $1\sigma$ confidence intervals.  These are not considered in the combination and the \CoMEnergy\ result is taken as the Run~1 result.

The BLUE combination requires the measured values and uncertainties of the parameters in question as well as the correlations between them.  
These are provided by the fits separately in the \CoMEnergySeven\ and \CoMEnergy\ measurements.   The statistical correlation between these two measurements is zero as the events are different.  The correlations of the systematic uncertainties between the two measurements are estimated by splitting the uncertainty into several categories.  

The  trigger efficiency is included as a systematic uncertainty only in the \CoMEnergySeven\ measurement, so there is no correlation with the \CoMEnergy\ measurement.  Similarly, the systematic uncertainties arising from the $\Lambda_b \ra J/\psi p K^{-}$ background, and the choice of \psubt\ bins and mass sidebands in the modelling of background angles, are included as systematic uncertainties only in the \CoMEnergy\ measurement so there is no correlation with the  \CoMEnergySeven\ measurement.  In both the \CoMEnergySeven\ and \CoMEnergy\ results, a systematic uncertainty is assigned to the inner detector alignment and $B_d$ contribution.  The inner detector alignment systematic uncertainties are highly correlated and small.  The assumed correlation between these systematics made no difference to the final combined result and was set to 100\%.  For the $B_d$ contribution, while the systematic uncertainty tests are different, they are both performed to account for an imprecise knowledge of the $B_d$ contribution and are therefore assumed to be 100\%.  The tagging, acceptance and fit model uncertainties are quoted for both  \CoMEnergySeven\  and  \CoMEnergy.  For the fit model, there are several different model variations each with their own uncertainty.  For each year, these are summed in quadrature to produce a single fit model systematic uncertainty.

The tagging, acceptance and fit model systematic uncertainties are each assigned a variable ($\rho_i$, where $i = \rm tag, acc, mod$) corresponding to the correlation between the \CoMEnergySeven\ and \CoMEnergy\ results.  Several different combinations were tried with different values of $\rho_i = 0, 0.25, 0.5, 0.75, 1.0$.  The acceptance systematic uncertainty is small and therefore regardless of what value of $\rho_{\rm acc}$ is chosen the combination stays the same.  For the \CoMEnergy\ measurement, electron tagging is added, therefore the systematic uncertainty is not $100\%$ correlated.  For $\rho_{\rm tag} = 0.25, 0.5, 0.75$ there is negligible difference between the results.  The fit model was changed between the \CoMEnergySeven\ and  \CoMEnergy\ measurement, the most significant change is that the mass uncertainty modelling was removed and the event-by-event Gaussian error distribution was replaced with a sum of three Gaussian distributions.  It would be incorrect to estimate the correlation as $100\%$ and there is negligible difference between the results for $\rho_{\rm{mod}} = 0.25, 0.5, 0.75$.  

The combined results for the fit parameters and their uncertainties for Run~1 are given in Table \ref{tab:FitResultsATLAS}. Due to the negative correlation between \Gs\ and \DGs, and the change in the value of \DGs\ between the \CoMEnergySeven\ and \CoMEnergy\ results, the combined value of \Gs\ is less than either individual result.
The Run~1 likelihood contours in the \phis--\DGs\ plane are shown in Figure \ref{fig:Contour}. They agree with the Standard Model predictions.

\begin{table}
\begin{center}
\begin{tabular}{c| c|c|c| c|c|c| c|c|c   }
\hline

 & \multicolumn{3}{c|}{  \CoMEnergy\ data } & \multicolumn{3}{c|}{  \CoMEnergySeven\ data} & \multicolumn{3}{c}{ Run1 combined}\\
Par & Value & Stat & Syst & Value & Stat & Syst & Value & Stat & Syst \\
\hline\hline

 \phis [rad] & \pSfit & \pSstat & \pSsyst   & \pSfiteleven & \pSstateleven & \pSsysteleven & \pSfitCombi & \pSstatCombi & \pSsystCombi  \\
 
 \DGs [ps$^{-1}$] & \DGfit & \DGstat & \DGsyst & \DGfiteleven & \DGstateleven & \DGsysteleven & \DGfitCombi & \DGstatCombi & \DGsystCombi\\
 
 \Gs [ps$^{-1}$] & \GSfit & \GSstat & \GSsyst & \GSfiteleven & \GSstateleven & \GSsysteleven & \GSfitCombi & \GSstatCombi & \GSsystCombi \\
 
  $|A_{\parallel}(0)|^2$ & \Apasqfit & \Apasqstat & \Apasqsyst & \Apasqfiteleven & \Apasqstateleven & \Apasqsysteleven & \ApasqfitCombi & \ApasqstatCombi & \ApasqsystCombi  \\
  
  $|A_{0}(0)|^2$ & \Azesqfit  & \Azesqstat & \Azesqsyst & \Azesqfiteleven  & \Azesqstateleven & \Azesqsysteleven & \AzesqfitCombi  & \AzesqstatCombi & \AzesqsystCombi \\
  
$|A_{S}|^2$ & \ASsqfit & \ASsqstat & \ASsqsyst & \ASsqfiteleven & \ASsqstateleven & \ASsqsysteleven & \ASsqfitCombi & \ASsqstatCombi  &\ASsqsystCombi \\ 

$\delta_\perp$ [rad] & \Deltperpfit  & \Deltperpstat & \Deltperpsyst &\Deltperpfiteleven  & \Deltperpstateleven   & \Deltperpsysteleven & \DeltperpfitCombi  & \DeltperpstatCombi & \DeltperpsystCombi \\[3pt]

$\delta_{\parallel}$ [rad] & \Deltparafit & \Deltparastat &\Deltparasyst &  \multicolumn{2}{c|}{\Deltparafiteleven} & \Deltparasysteleven & \DeltparafitCombi & \DeltparastatCombi &\DeltparasystCombi \\[3pt]

$\delta_{\perp} -\delta_{S}$ [rad] & \DeltPSfit & \DeltSstat & \DeltPSsyst  & \multicolumn{2}{c|}{\DeltPSfiteleven } & \DeltPSsysteleven  & \DeltPSfitCombi  & \DeltPSstatCombi & \DeltPSsystCombi  \\

\hline
\end{tabular}
\end{center}
\caption{ Current measurement using data from \CoMEnergy\ $pp$ collisions, the previous measurement using  data taken at centre of mass energy of \CoMEnergySeven\, and the values for the parameters of the two measurements, statistically combined. }
 \label{tab:FitResultsATLAS}
\end{table}

\section{Summary}
A measurement of the time-dependent $CP$ asymmetry parameters in \Bsto\ decays from a \ilumi\ data sample of $pp$ collisions collected with the ATLAS detector during the \CoMEnergy\ LHC run is presented.  The values from the \CoMEnergy\ analysis are consistent with those obtained in the previous analysis using \CoMEnergySeven\  ATLAS data \cite{tagCPV2011}. The two measurements are statistically combined  leading to the following results:
\begin{eqnarray*}
\phis & = & \pSfitCombi \pm \pSstatCombi~\rm{(stat.)}\pm \pSsystCombi~\rm{(syst.)}~rad  \\
\DGs & = & \DGfitCombi \pm \DGstatCombi~\rm{(stat.)}\pm \DGsystCombi~\rm{(syst.)~ps}^{-1} \\
\Gs & = & \GSfitCombi \pm \GSstatCombi ~\rm{(stat.)}\pm \GSsystCombi ~\rm{(syst.)~ps}^{-1}\\
|A_{\parallel}(0)|^2 & = & \ApasqfitCombi \pm \ApasqstatCombi~\rm{(stat.)}\pm \ApasqsystCombi~\rm{(syst.)}\\
|A_{0}(0)|^2 & = & \AzesqfitCombi \pm \AzesqstatCombi~\rm{(stat.)}\pm \AzesqsystCombi~\rm{(syst.)}\\
|A_{S}(0)|^2    & = &  \ASsqfitCombi \pm \ASsqstatCombi~\rm{(stat.)} \pm \ASsqsystCombi~\rm{(syst.)}\\ 
\delta_\perp & = & \DeltperpfitCombi \pm \DeltperpstatCombi ~\rm{(stat.)}\pm \ \DeltperpsystCombi~\rm{(syst.)}~rad \\
\delta_{\parallel}  & = & \DeltparafitCombi \pm \DeltparastatCombi ~\rm{(stat.)}\pm \ \DeltparasystCombi~\rm{(syst.)}~rad \\
 \delta_{\perp}-\delta_S & = &  \DeltPSfitCombi \pm \DeltPSstatCombi~\rm{(stat.)} \pm \DeltPSsystCombi~\rm{(syst.)}~rad.  
\end{eqnarray*}
The ATLAS Run 1 results for the \Bst \ decay are consistent with the SM.

%% file: acknowledgements/Acknowledgements.tex

We thank CERN for the very successful operation of the LHC, as well as the
support staff from our institutions without whom ATLAS could not be
operated efficiently.

We acknowledge the support of ANPCyT, Argentina; YerPhI, Armenia; ARC, Australia; BMWFW and FWF, Austria; ANAS, Azerbaijan; SSTC, Belarus; CNPq and FAPESP, Brazil; NSERC, NRC and CFI, Canada; CERN; CONICYT, Chile; CAS, MOST and NSFC, China; COLCIENCIAS, Colombia; MSMT CR, MPO CR and VSC CR, Czech Republic; DNRF and DNSRC, Denmark; IN2P3-CNRS, CEA-DSM/IRFU, France; GNSF, Georgia; BMBF, HGF, and MPG, Germany; GSRT, Greece; RGC, Hong Kong SAR, China; ISF, I-CORE and Benoziyo Center, Israel; INFN, Italy; MEXT and JSPS, Japan; CNRST, Morocco; FOM and NWO, Netherlands; RCN, Norway; MNiSW and NCN, Poland; FCT, Portugal; MNE/IFA, Romania; MES of Russia and NRC KI, Russian Federation; JINR; MESTD, Serbia; MSSR, Slovakia; ARRS and MIZ\v{S}, Slovenia; DST/NRF, South Africa; MINECO, Spain; SRC and Wallenberg Foundation, Sweden; SERI, SNSF and Cantons of Bern and Geneva, Switzerland; MOST, Taiwan; TAEK, Turkey; STFC, United Kingdom; DOE and NSF, United States of America. In addition, individual groups and members have received support from BCKDF, the Canada Council, CANARIE, CRC, Compute Canada, FQRNT, and the Ontario Innovation Trust, Canada; EPLANET, ERC, FP7, Horizon 2020 and Marie Sk{\l}odowska-Curie Actions, European Union; Investissements d'Avenir Labex and Idex, ANR, R{\'e}gion Auvergne and Fondation Partager le Savoir, France; DFG and AvH Foundation, Germany; Herakleitos, Thales and Aristeia programmes co-financed by EU-ESF and the Greek NSRF; BSF, GIF and Minerva, Israel; BRF, Norway; Generalitat de Catalunya, Generalitat Valenciana, Spain; the Royal Society and Leverhulme Trust, United Kingdom.

The crucial computing support from all WLCG partners is acknowledged gratefully, in particular from CERN, the ATLAS Tier-1 facilities at TRIUMF (Canada), NDGF (Denmark, Norway, Sweden), CC-IN2P3 (France), KIT/GridKA (Germany), INFN-CNAF (Italy), NL-T1 (Netherlands), PIC (Spain), ASGC (Taiwan), RAL (UK) and BNL (USA), the Tier-2 facilities worldwide and large non-WLCG resource providers. Major contributors of computing resources are listed in Ref.~\cite{ATL-GEN-PUB-2016-002}.

%% file: atlas_authlist.tex
\begin{flushleft}
{\Large The ATLAS Collaboration}

\bigskip

G.~Aad$^{\rm 85}$,
B.~Abbott$^{\rm 113}$,
J.~Abdallah$^{\rm 151}$,
O.~Abdinov$^{\rm 11}$,
R.~Aben$^{\rm 107}$,
M.~Abolins$^{\rm 90}$,
O.S.~AbouZeid$^{\rm 158}$,
H.~Abramowicz$^{\rm 153}$,
H.~Abreu$^{\rm 152}$,
R.~Abreu$^{\rm 116}$
% ...
\bigskip
\\
% ...
$^{11}$ Institute of Physics, Azerbaijan Academy of Sciences, Baku, Azerbaijan\\
% ...
$^{85}$ CPPM, Aix-Marseille Universit{\'e} and CNRS/IN2P3, Marseille, France\\
% ...
$^{90}$ Department of Physics and Astronomy, Michigan State University, East Lansing MI, United States of America\\
% ...
$^{107}$ Nikhef National Institute for Subatomic Physics and University of Amsterdam, Amsterdam, Netherlands\\
% ...
$^{113}$ Homer L. Dodge Department of Physics and Astronomy, University of Oklahoma, Norman OK, United States of America\\
% ...
$^{116}$ Center for High Energy Physics, University of Oregon, Eugene OR, United States of America\\
% ...
$^{151}$ Institute of Physics, Academia Sinica, Taipei, Taiwan\\
$^{152}$ Department of Physics, Technion: Israel Institute of Technology, Haifa, Israel\\
$^{153}$ Raymond and Beverly Sackler School of Physics and Astronomy, Tel Aviv University, Tel Aviv, Israel\\
% ...
$^{158}$ Department of Physics, University of Toronto, Toronto ON, Canada\\
% ...
\end{flushleft}
